%% file: 0_baryon_DES.tex
\documentclass[usenatbib]{mnras}    

\usepackage[T1]{fontenc}
\usepackage{ae,aecompl}

\usepackage{graphicx}	
\usepackage{amsmath}	
\usepackage{amssymb}	

\usepackage{color}
\usepackage{bm}
\usepackage{bbm}
\usepackage{makecell}
\usepackage{multirow}
\usepackage{hyperref}

\usepackage{lscape}
\usepackage{rotating}
\usepackage{cancel}

\usepackage{newtxtext,newtxmath}

\usepackage{eso-pic}

\AddToShipoutPictureBG*{%
  \AtPageUpperLeft{%
    \hspace{0.75\paperwidth}%
    \raisebox{-3.5\baselineskip}{%
      \makebox[0pt][l]{\textnormal{DES-2020-0557}}
}}}%

\AddToShipoutPictureBG*{%
  \AtPageUpperLeft{%
    \hspace{0.75\paperwidth}%
    \raisebox{-4.5\baselineskip}{%
      \makebox[0pt][l]{\textnormal{FERMILAB-PUB-20-286-AE}}
}}}%

\definecolor{green2}{rgb}{0.13, 0.55, 0.13}
\definecolor{gray2}{rgb}{0.6, 0.6, 0.6}
\definecolor{orange}{rgb}{1.0, 0.49, 0}
\definecolor{blue2}{rgb}{0.25,0.5,1}
\definecolor{fireenginered}{rgb}{0.81, 0.09, 0.13}

\input{def.tex}

\defcitealias{Eifler15}{E15}
\defcitealias{Huang19}{H19}

\title[Baryonic Physics in the Universe]
{Dark Energy Survey Year 1 Results: Constraining Baryonic Physics in the Universe}

\input{author_list.tex}

\date{Accepted XXX. Received YYY; in original form ZZZ}

\pubyear{2020}

\begin{document}
\label{firstpage}
\pagerange{\pageref{firstpage}--\pageref{lastpage}}
\maketitle

\begin{abstract} 
Measurements of large-scale structure are interpreted using theoretical predictions for the matter distribution, including potential impacts of baryonic physics. 
We constrain the feedback strength of baryons jointly with cosmology using weak lensing and galaxy clustering observables (3$\times$2pt) of Dark Energy Survey (DES) Year 1 data in combination with external information from baryon acoustic oscillations (BAO) and Planck cosmic microwave background polarization.
Our baryon modeling is informed by a set of hydrodynamical simulations that span a variety of baryon scenarios; we span this space via a Principal Component (PC) analysis of the summary statistics extracted from these simulations. 
We show that at the level of DES Y1 constraining power, one PC is sufficient to describe the variation of baryonic effects in the observables, and the first PC amplitude ($Q_1$) generally reflects the strength of baryon feedback.
With the upper limit of $Q_1$ prior being bound by the Illustris feedback scenarios, we reach $\sim 20\%$ improvement in the constraint of $S_8=\sigma_8(\Omega_{\rm m}/0.3)^{0.5}=0.788^{+0.018}_{-0.021}$ compared to the original DES 3$\times$2pt analysis.  This gain is driven by the inclusion of small-scale cosmic shear information down to $2.5\arcmin$, which was excluded in previous DES analyses that did not model baryonic physics. 
We obtain $S_8=0.781^{+0.014}_{-0.015}$ for the combined DES Y1+Planck EE+BAO analysis with a non-informative $Q_1$ prior.
In terms of the baryon constraints, we measure $Q_1=1.14^{+2.20}_{-2.80}$ for DES Y1 only and $Q_1=1.42^{+1.63}_{-1.48}$ for DESY1+Planck EE+BAO, allowing us to exclude one of the most extreme AGN feedback hydrodynamical scenario at more than $2 \sigma$. 
\end{abstract}




\section{Introduction}
\label{sec:intro}
Understanding the composition and evolution of our Universe has been a central science endeavor in the astronomical community. Ongoing wide-field imaging surveys such as the Dark Energy Survey \citep[DES\footnote{www.darkenergysurvey.org/},][]{Krause17, Troxel18, DES18,DES5x2}, the Kilo-Degree Survey \citep[KiDS\footnote{http://www.astro-wise.org/projects/KIDS/},][]{vanUitert18,Kuijken19,Hildebrandt20}, and the Hyper Suprime Cam Subaru Strategic Program \citep[HSC\footnote{http://www.naoj.org/Projects/HSC/HSCProject.html},][]{mmh18,Hikage19,Hamana20} have collected a wealth of cosmological data over the past years that can be used to explore fundamental questions such as the underlying physics of cosmic acceleration, the mass and number of neutrino species, and the interplay of dark and luminous matter.  

The cosmological information is bound to increase significantly in the near future with analysis of the full DES, KiDS, and HSC datasets, and even more so in the early 2020s with the advent of Stage IV surveys such as the Vera C.\ Rubin Observatory Legacy Survey of Space and Time \citep[LSST\footnote{https://www.lsst.org/},][]{LSST19}, Euclid\footnote{https://sci.esa.int/web/euclid} \citep{laa11}, the Spectro-Photometer for the History of the Universe, Epoch of Reionization, and Ices Explorer \citep[SPHEREx\footnote{http://spherex.caltech.edu/},][]{dba14}, and the Nancy Grace Roman Space Telescope \citep[WFIRST\footnote{https://wfirst.gsfc.nasa.gov/},][]{sgb15,esk20,emk20}. 

The increased cosmological information encoded in these datasets will require a new level in accuracy of modeling cosmological observables. One of the fundamental quantities for making theoretical predictions is the matter power spectrum $P_{\delta}(k, z)$, which quantifies the amount of matter clustering at the second-order level and its evolution as a function of time. 
Previous studies have estimated that $P_{\delta}(k, z)$ needs to be predicted to $\sim 1\%$ level out to $k \lesssim 10$ $h$Mpc$^{-1}$ for the future era of Stage IV cosmological experiments \citep[e.g.,][]{Huterer05, Eifler11, Hearin12}. To quantify the nonlinear evolution of the density field at the required precision, significant computational resources have been devoted to building power spectrum emulators with N-body dark-matter-only (DMO) simulations \citep[e.g.,][]{Heitmann10, Heitmann14, DeRose19}.
However, baryonic effects such as feedback and cooling mechanisms redistribute  matter, causing uncertainties in $P_{\delta}(k, z)$ at the level of tens of per cent \citep[e.g.,][]{vanDaalen11,Chisari18,vanDaalen20} for $k \gtrsim 5$ $h$Mpc$^{-1}$.

Adopting mitigation schemes to account for uncertainties of baryons is crucial to assure the robustness of cosmological analyses. The most straightforward way is to exclude data points for which the fractional contributions from potential systematic uncertainties are non-negligible given the covered model flexibility. For the DES Y1 cosmic shear analysis, conservative scale cuts are applied to ensure the level of baryon contamination to be within 2\% \citep{Troxel18}.

Methods have been proposed to reduce the sensitivity to small scales in the data. 
By cutting the most extreme peaks in the density fields, the derived summary statistics become less sensitive to the non-linear regime, as proposed in the peak clipping technique \citep{Simpson11,Simpson13,Giblin18}.
By designing special weighting functions to filter out the contributions of small-scale modes in observables of cosmic shear, the $k$-cut \citep{Taylor18} and $x$-cut \citep{Taylor20} cosmic shear methods provide new summary statistics with reduced sensitivity to baryonic effects on the matter power spectrum. 
Also, the \textsc{COSEBIs} \citep{Schneider10} method is designed to separate E/B modes from $\xi_{\pm}$ in a finite angular interval, which makes its summary statistics less sensitive to small scale physical effects compared with $\xi_{\pm}$ given a fixed angular range \citep{Asgari20}.

However, including small-scale information with models for baryonic effects not only provides the potential to increase the statistical power of constraints on cosmology, but also offers a mechanism to quantify the effects of baryons on the matter power spectrum using real data.
A number of methods have been proposed to model baryonic effects (see \citealt{Chisari19} for a review). 
One class of methods is to employ the halo model \citep{Peacock00, Seljak00}, based on the principle that the main contribution of baryons is to modify halo density profiles in the one-halo regime (see e.g. \citealt{Rudd08, Velliscig14, Mummery17}).
Within the NFW (Navarro-Frenk-White, \citealt{Navarro96}) profile, a straightforward option is to vary parameters related to the halo concentration to perform baryon marginalization \citep{Zentner08, Zentner13}.
Besides the degree of freedom provided via halo concentration, extra parameters are added offering the complexity to account for the effect of halo bloating induced by baryonic feedback in \texttt{HMcode} \citep{Mead15, Mead16}, and for the inner halo core formation induced by the cooling effect of baryons \citep{Copeland18}. 
\texttt{HMcode} has been applied in several weak lensing analyses to mitigate baryonic effects, for example in data sets of CFHTLenS \citep{Joudaki17}, DES Science Verification \citep{MacCrann17}, KiDS-450 \citep{Hildebrandt17, Yoon20}, and DLS \citep{Yoon19}. Even more sophisticated halo model frameworks provide descriptions of the radial distributions of the stellar, gas, and dark matter components within haloes, and parametrize the baryonic effects in more physically motivated quantities \citep{Semboloni11, Semboloni13, Mohammed14, Schneider15, Debackere20}.

Another category of approaches to modeling baryonic effects is through empirical modeling, where the functional form of the fitting formula is calibrated based on hydrodynamical simulations. Parametric forms are designed with the flexibility to model the behavior of the power spectrum ratio between paired hydrodynamical and DMO simulations ($P_{\delta, {\rm hydro}}(k)$/$P_{\delta, {\rm DMO}}(k)$) for the Horizon-AGN hydro-simulation in \citet{Chisari18}, and for the nine scenarios in the OWLS simulation suites as detailed in \citet{HarnoisDeraps15}, which is adopted in HSC Y1 cosmic shear analysis to account for baryonic effects \citep{Hikage19}.
Recently \citet{vanDaalen20} derive a formulation which provides even wider applications for hydrodynamical scenarios accumulated over the past ten years.
Alternatively, \citet{Eifler15} proposed performing principal component analyses (PCA) using the cosmic shear model vectors extracted from the hydrodynamical simulations, and use a few dominant principal components (PCs) as a flexible basis to span the range of baryon uncertainties for the survey-specific summary statistic.

Going beyond modeling summary statistics, there are approaches focusing on  post-processing the output DMO simulations. The `baryonification' model contains prescriptions for the density profiles of the stellar, gas, and the redistributed DM components to correct the particle positions in DMO simulations so as to more accurately approximate what they would look like in the presence of baryons \citep{Schneider15, Schneider19, Arico20}. 
\citet{Dai18} propose using the potential gradient descent model to displace particles to improve the modeling of non-linear matter distribution.

In this paper, we aim to utilize the information from small-scale cosmic shear data to place constraints on the strength of baryonic effects, and to compare the results with existing hydrodynamical simulations. We will also explore the potential for achieving more precise cosmological constraints with the inclusion of small-scale data. 
We adopt the PCA baryon mitigation framework \citep{Eifler15} to perform our analyses. In \citet{Huang19}, we have validated and improved the performance of the PCA method using simulated analyses of cosmic shear mock data under an LSST-like survey configuration.
Here we delve into its application to the observational data of DES Y1, which includes  two-point correlations of cosmic shear, galaxy-galaxy lensing and galaxy clustering. With the ability to model small-scale cosmic shear, we push the cosmic shear observables down to $2.5\arcmin$ and perform a combined analysis with galaxy-galaxy lensing and galaxy clustering data (subjected to the original conservative Y1 scale cuts).

We begin with an overview of the data products, the theoretical modeling, and the analysis approaches in \S\ref{sec:theory}. We describe the design and validation of our pipeline in \S\ref{sec:pipeline}. We employ simulated likelihood analyses to understand and validate our pipeline performance, before we unblind and perform analyses of the real DES Y1 data.
We present our main cosmology results in \S\ref{sec:cosmology_constraint}, followed by our constraints on baryonic effects in \S\ref{sec:baryon_constraint}. Finally, we conclude in \S\ref{sec:summary}.

\section{Data, Theory, and Analysis}
\label{sec:theory}

\subsection{Data}
\label{subsec:data}

\subsubsection{Observational Data}
\label{subsec:DES_data}

In this work, we use the DES Y1 3$\times$2pt data vector\footnote{The publicly released 3$\times$2pt data vector 
and its associated covariance matrix, \texttt{2pt\_NG\_mcal\_1110.fits}, can be downloaded at 
\href{https://des.ncsa.illinois.edu/releases/y1a1/key-products}
{https://des.ncsa.illinois.edu/releases/y1a1/key-products}} which is computed using the \metacal \citep{Huff17, Sheldon17, Zuntz18} shape catalog as 
 the source sample for cosmic shear \citep{Troxel18}, and the redMaGiC \citep{Rozo16} sample as the lens 
population for galaxy-galaxy lensing \citep{Prat18} and galaxy clustering \citep{Elvin-Poole18} measurements. The photometric redshift measurement and calibration are described in \citet{Hoyle18, Gatti18, Davis17}.

The DES Y1 source galaxies are divided into four tomographic bins ranging from $z=0.2$ to $1.3$, resulting in 10 auto- and cross-correlations of cosmic shear for $\xi_{+}$ and $\xi_{-}$, respectively. The lens galaxies are placed in five tomographic bins ranging from $z=0.15$ to $0.9$, resulting in 20 tomographic cross-correlation bins between lens and source samples for galaxy-galaxy lensing, and 5 auto-correlations for galaxy clustering. Each of the correlation function statistics is measured using \texttt{treecorr} \citep{Jarvis04} in 20 log-spaced bins of angular separation $ 2.5' < \theta < 250'$. 

Conservative scale cuts are applied to the raw 3$\times$2pt data vector in the original DES Y1 key cosmological analysis to avoid biases due to modeling uncertainties on small scales \citep{DES18}. 

For cosmic shear, scale cuts are determined by contaminating the $\xi^{\pm}$ model vector according to the OWLS-AGN scenario \citep{Schaye10}, which has the same baryonic feature as the cosmo-OWLS AGN scenario shown in the red curves in Fig.~\ref{fig:Pk_Ratio}, and removing data points that have a  fractional contribution of baryons exceeding 2\% \citep{Troxel18}. For galaxy-galaxy lensing and galaxy clustering, the scale cuts are defined using a specific comoving scales of $(R_{\rm ggl}, R_{\rm clustering})=(12, 8)$ Mpc $h^{-1}$ to avoid parameter biases due to non-linear biasing or non-locality of $\gamma_{\rm t}$, and converted to their corresponding angular scales in each tomographic bin \citep{Krause17}.
After scale cuts are applied, there are a total of 457 elements for the fiducial DES Y1 3$\times$2pt cosmological analysis \citep{DES18}.

In this analysis, we utilize the DES Y1 cosmic shear correlation function measurements down to scales of 2.5$\arcmin$. Together with the galaxy-galaxy lensing and galaxy clustering measurements (subjected to the original DES Y1 scale cuts), our extended 3$\times$2pt data vector has a total of 630 data points (400 elements for cosmic shear, 176 elements for galaxy-galaxy lensing and 54 elements for galaxy clustering).

\subsubsection{Hydrodynamical Simulation Data and Power Spectrum}
\label{subsec:hydrosims}

In order to build baryon mitigation models with sufficient flexibility, we rely on a large variety of hydrodynamical simulations: MassiveBlack-II (MB2, \citealt{Khandai15, Tenneti15b}), Horizon-AGN \citep{Dubois14}, Eagle \citep{Schaye15}, Illustris \citep{Vogelsberger14,Genel14}, IllustrisTNG \citep{Springel18,Pillepich18b,Naiman18,Marinacci18,Nelson18}, three cosmo-OWLS simulations (cOWLS, \citealt{LeBrun14}) with their minimum active galactic nucleus (AGN) heating temperatures $\Delta T_{\rm heat}$ being set at $10^{8.0}$, $10^{8.5}$, $10^{8.7}$, and three BAHAMAS scenarios \citep{McCarthy17} with their $\Delta T_{\rm heat} = 10^{7.6}$, $10^{7.8}$, $10^{8.0}$. Here $\Delta T_{\rm heat}$ is the most dominant subgrid physical pararmter controlling the strength of AGN feedback in the cosmo-OWLS and the BAHAMAS simulation sets. Black holes are storing their feedback energy until it is large enough to heat the a certain number of surrounding particles by $\Delta T_{\rm heat}$. 


\begin{figure*}
\begin{center}
\includegraphics[width=1.0\textwidth]{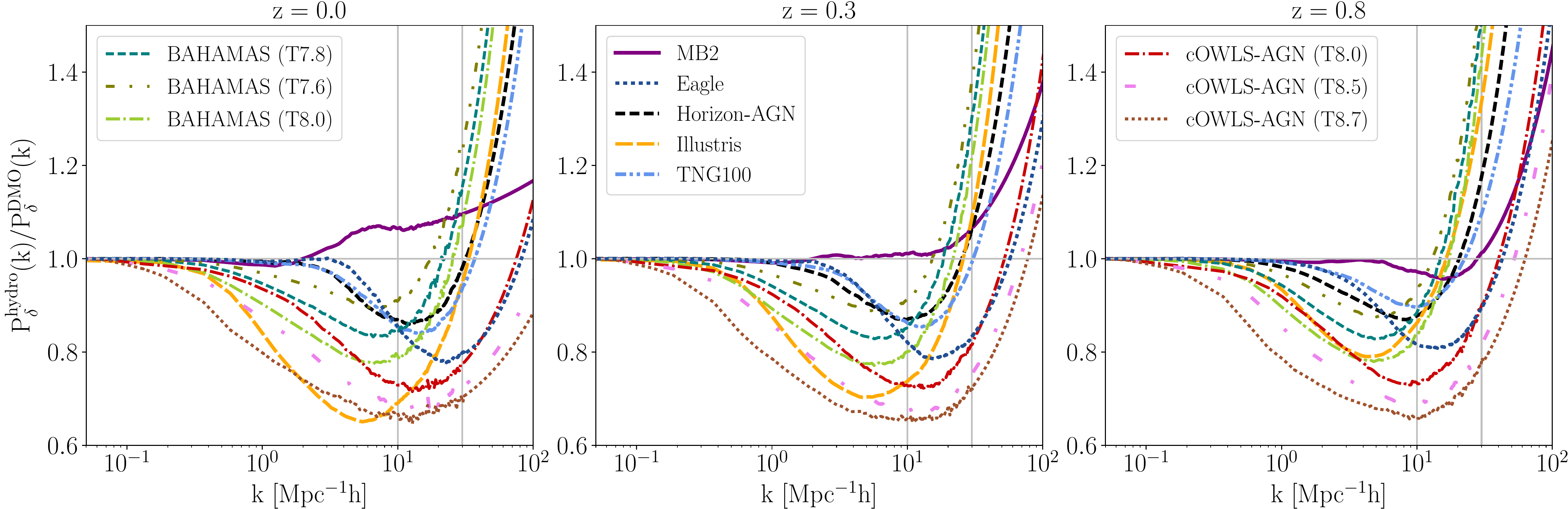}
\caption{Baryonic effects on the 3D matter power spectrum at different redshifts. We plot the power spectrum ratio for 11 hydrodynamical simulations with respect to their corresponding DMO simulation setting at the same initial condition: IllustrisTNG, MB2, Eagle, Horizon-AGN, Illustris, the cosmo-OWLS sets, and the BAHAMAS sets, at redshifts 0.0, 0.3, 0.8.
The gray vertical lines delineate regions where the data points come from direct measurement ($k < 30$ $h$ Mpc$^{-1}$) and from extrapolation ($k > 30$ $h$ Mpc$^{-1}$) with a quadratic spline fit based on data points at $k \in [10, 30]$ $h$ Mpc$^{-1}$.}
\label{fig:Pk_Ratio}
\end{center}
\end{figure*}

Figure~\ref{fig:Pk_Ratio} shows the effects of baryonic physics on the 3D matter power spectra for different hydrodynamical simulations, displayed as the ratio of these power spectra with respect to the power spectra for the corresponding dark matter only (DMO) simulations with the same initial conditions. On small scales, the effects of baryons show large variations, and have different redshift evolution histories across simulations. On large scales, we expect the power spectrum ratios to converge to unity because of diminishing baryonic effects, and because of the cosmic variance fluctuations being canceled when taking ratios of power spectra for pairs of simulations with identical initial conditions. 
In Appendix~B of \citealt{Huang19} (hereafter \citetalias{Huang19}), we have discussed the convergence of power spectrum ratios in detail and provide an upper limit for their uncertainties due to cosmic variance.

We have used the power spectrum ratio for MB2, Illustris, Eagle from \citetalias{Huang19}. We extracted power spectrum measurements from the publicly released IllustrisTNG100 snapshot data \citep{Nelson19} and added the corresponding baryonic scenario to our power spectrum library. The Horizon-AGN $P_\delta(k)$ data are computed in \citet{Chisari18}. The cosmo-OWLS and BAHAMAS $P_\delta(k)$ sets are taken from the power spectra library released by \citet{vanDaalen20}. Specifically, as listed in Table 1 of \citet{vanDaalen20}, for the cosmo-OWLS baryonic scenario sets, we use the $P_\delta(k)$ data from files: \begin{itemize}
\item \textit{AGN\_Mseed800\_WMAP7\_L100N512},
\item \textit{AGN\_Mseed800\_Theat 8p5\_WMAP7\_L100N512}, \item \textit{AGN\_Mseed800\_Theat 8p7\_WMAP7\_L100N512}, \end{itemize}
for the BAHAMAS sets, we use files: 
\begin{itemize}
\item \textit{AGN\_CALIB\_nu0\_WMAP9\_L400N1024}, 
\item \textit{AGN\_CALIB\_Theat\_7p6\_nu0\_WMAP9\_L400N1024}, 
\item \textit{AGN\_CALIB\_Theat\_8p0\_nu0\_WMAP9\_L400N1024}. 
\end{itemize}

We make a slight adjustment to the power spectrum ratios. At larger scales, the raw $P_\delta(k)$ ratios for Horizon-AGN and cosmo-OWLS are observed to have subtle ($ \lesssim 1\%$) excesses above unity toward large scales (e.g. see Fig.~5 of \citealt{Chisari18}). 
As discussed in Appendix~B of \citet{vanDaalen20}, this large-scale excess of power originates from details of the simulation setup between pairs of hydrodynamical and DMO simulations, for which their transfer functions and the number of particles often differ. Given that this sub-percent level offset is due to artifacts, we correct for this power mismatch by re-scaling the DMO power spectra using the linear growth factor, such that the ratio between $P_\delta^{\rm hydro}$ and $P_\delta^{\rm DMO}$ asymptotically approaches one on large scales. 

On scales above $k > 30$ Mpc$^{-1}$ $h$, we perform extrapolation by fitting a quadratic spline curve to data points at $k \in [10, 30]$ Mpc$^{-1}$ $h$ to capture the power boosting from the effect of cooling. As discussed in Appendix B of \citetalias{Huang19}, we argue that our extrapolation approach more accurately captures cooling effects compared to simply adopting the raw ratio as computed from the simulations. This is supported by comparing both methods to power spectrum ratios derived from higher resolution simulations.

\subsubsection{Mock Data Vectors}
\label{subsec:mock_data}

In order to validate our baryon mitigation pipeline, we generate three mock data vectors to conduct simulated likelihood analyses: a pure theoretical data vector derived from our analysis pipeline (\textsc{CosmoLike}) with the fiducial parameters shown in Table~\ref{tb:params} (we refer to this mock data vector as the DMO scenario hereafter), and two baryon-contaminated mock data vectors based on the Illustris and Eagle scenarios. Throughout this work, when conducting a simulated analysis with a specific baryon-contaminated mock data vector, we avoid using this specific baryonic scenario as input to the construction of our baryon mitigation model. Further details of the simulated likelihood analyses are found in \S\ref{subsec:PCA_overview}.

We derive the baryon-contaminated data vectors at a specific cosmology $\pco$ using the underlying hydrodynamical power spectrum defined as 
\begin{equation}
\label{eq:Pk_ratio}
P_{\delta}^{\rm hydro}(k, z\ |\ \pco) = \frac{  P_{\delta}^{\rm hydro, sim}(k, z\ |\ \pcosim)  }{  P_{\delta}^{\rm DMO, sim}(k, z\ |\ \pcosim)  } P_{\delta}^{\rm theory}(k, z\ |\ \pco) \ .
\end{equation} 
The ratio terms $\frac{  P_{\delta}^{\rm hydro, sim}(k, z\ |\ \pcosim)  }{  P_{\delta}^{\rm DMO, sim}(k, z\ |\ \pcosim) }$ are computed from interpolating the power spectrum ratio table constructed from various simulations snapshots from $z=0\sim 3.5$.\footnote{The power spectra ratio data are available at \href{https://github.com/hungjinh/baryon-power-spectra}{https://github.com/hungjinh/baryon-power-spectra} and from the \citet{vanDaalen20} \href{http://powerlib.strw.leidenuniv.nl}{data release}.} Some of the selected snapshots are visualized in Fig.~\ref{fig:Pk_Ratio}.

When using Eq.~\eqref{eq:Pk_ratio}, we implicitly assume that baryonic effects and cosmology are independent. That is, we fix the ratio of $P_{\delta}(k, z)$ for each baryonic scenario, while the cosmological dependence is propagated through the theoretical power spectrum $P_{\delta}^{\rm theory}(k, z\ |\ \pco)$. The expression of $P_{\delta}^{\rm hydro}(k, z\ |\ \pco)$ computed as in Eq.~\eqref{eq:Pk_ratio} is then passed into the \textsc{CosmoLike} package to derive the baryon-contaminated data vectors (\S\ref{subsec:model}). 

The dependence of the baryonic suppression of the power spectrum with cosmological parameters is explored in detail in \citet{Schneider20}. Based on the parametrization of their baryon correction model, the derived power spectrum ratios are largely independent of individual cosmological parameters, but they are related to the cosmic baryon fraction, $f_{\rm b}=\Omegab/\Omegam$ (see their Fig.~2); when varying $f_{\rm b}$ from 0.16 to 0.2, the power spectrum ratios are further suppressed by $\sim$5\% (10\%) for $k \leqslant 3$ $h$ Mpc$^{-1}$ ($k \leqslant$ 10 $h$ Mpc$^{-1}$). Here we use $k \sim$ 3 $h$ Mpc$^{-1}$ as a reference, because this roughly corresponds to the effective scale to which we are sensitive given our small-scale  cut at 2.5\arcmin, given the lensing kernel peak at $z\sim0.5$ for DES Y1. Similarily, based on a fixed sets of BAHAMAS runs, but varying cosmologies from WMAP 2009 ($f_{\rm b}\sim0.17$, \citealt{Hinshaw13}), Planck 2013 ($f_{\rm b}\sim0.15$, \citealt{Planck14}), to Planck 2015 ($f_{\rm b}\sim0.16$,  \citealt{Planck16}), \citet{vanDaalen20} showed that the power spectrum ratios vary $\lesssim$ 2\% (4\%) for $k \leqslant 3$ $h$ Mpc$^{-1}$ ($k \leqslant$ 10 $h$ Mpc$^{-1}$). 
We note that the interaction between baryonic physics and cosmological parameters is a subdominent effect given the constraining power of DES Y1. According to \citet{Schneider20}, ignoring the coupling of baryon suppression with $f_{\rm b}$ is a valid approximation even for future stage IV weak lensing surveys (see their Fig.~10).



\subsection{Model}
\label{subsec:model}

We use the \textsc{CosmoLike} package \citep{Krause17b}, one of the pipelines for DES cosmological inference, to perform the theoretical modeling of the 3$\times$2pt data vectors. The linear DMO power spectrum is generated at each cosmology using \textsc{class} \citep{Blas11}, with nonlinear corrections  derived from the \citet{Takahashi12} version of \textsc{Halofit}. 
Throughout this work, we consider a flat \LCDM cosmological model with six free parameters, $\pco=\{\Omegam,\ A_{\rm s},\ \Omegab,\ \ns,\ \Omega_{\nu} h^2,\ h\}$ in addition to the considered systematics parameters. The complete list of all parameters and their priors is given in Table~\ref{tb:params}.

Below we briefly summarize the theoretical modeling of the three types of two-point correlation functions and their associated systematic effects. 

\subsubsection{Cosmic Shear $\xi_{\pm}(\theta)$}

The real-space cosmic shear correlation function in tomographic bins $i$, $j$ is modeled as 
\begin{equation}
\label{eq:xi}
\xi^{ij}_{\pm}(\theta) = (1+ m^i) (1+ m^j)\ \frac{1}{2\pi} \int {\rm d} \ell \ \ell J_{0/4}(\ell \theta) C^{ij}_{\gamma \gamma} (\ell) \ .
\end{equation} Here $J_0$ and $J_4$ are Bessel functions of the first kind. The $m^i$ are multiplicative factors, one for each tomographic bin, that account for shear calibration bias \citep{Heymans06, Huterer06}. $C^{ij}_{\gamma \gamma} (\ell)$ is the detected shear-shear power spectrum, which contains the real lensing signal due to gravity (GG) as well as the contamination  due to intrinsic alignment (II, GI, IG terms)
\begin{equation}
\label{eq:Crr}
C^{ij}_{\gamma \gamma} (\ell) = C^{ij}_{\rm GG} (\ell) + C^{ij}_{\rm II} (\ell) + C^{ij}_{\rm GI} (\ell) + C^{ij}_{\rm IG} (\ell)\ .
\end{equation}

Adopting the Limber approximation and the flat Universe assumption (these modeling assumptions are demonstrated to be sufficient for Y1, see \citealt{Fang20}) the real lensing contribution can be computed as
\begin{equation}
\label{eq:CGG}
C^{ij}_{\rm GG} (\ell) = \int^{\chi_{\rm h}}_{0} {\rm d}\chi_{\rm l} \frac{g^i(\chi_{\rm l})g^j(\chi_{\rm l})}{\chi_{\rm l}^2} P_{\delta}(k=\frac{\ell}{\chi_{\rm l}}, \chi_{\rm l})\ ,
\end{equation}
where $\chi_{\rm l}$ is the comoving distance for the matter distribution (lens) along the line of sight, and $\chi_{\rm h}$ is the comoving horizon distance. The lensing kernel in the $i$-th tomographic interval is 
\begin{equation}
\label{eq:gi}
g^i(\chi_{\rm l}) = \frac{3}{2} \frac{H_0^2 \Omegam}{c^2} \frac{\chi_{\rm l}}{a(\chi_{\rm l})} \int^{\chi_{\rm h}}_{\chi_{\rm l}} {\rm d} \chi_{\rm s} n^i_{\rm s}(\chi_{\rm s}) \frac{\chi_{\rm s}-\chi_{\rm l}}{\chi_{\rm s}}\ ,
\end{equation}
with $n^i_{\rm s}(\chi_{\rm s})$ being the probability density function (pdf) for the redshift distribution of source galaxies in tomographic bin $i$, defined such that $n^i_{\rm s}(\chi_{\rm s}) {\rm d}\chi_{\rm s} = n^i_{\rm s}(z) {\rm d}z$, which is normalized to unity. 

For the intrinsic alignment (IA) contamination, we compute the intrinsic-intrinsic shape correlation due to the local tidal gravitational field on pairs of source galaxies as, 
\begin{equation}
\label{eq:CII}
C^{ij}_{\rm II} (\ell) = \int^{\chi_{\rm h}}_{0} {\rm d}\chi_{\rm s} \frac{n_{\rm s}^i(\chi_{\rm s})n^j_{\rm s}(\chi_{\rm s})}{\chi_{\rm s}^2} P_{\rm II}(k=\frac{\ell}{\chi_{\rm s}}, \chi_{\rm s})\ .
\end{equation}
The lensing shear-intrinsic shape correlations for pairs of galaxies where the foreground one is tidally torqued and the background one is sheared by the same gravitational field reads,  
\begin{equation}
\label{eq:CGI}
C^{ij}_{\rm GI} (\ell) + C^{ij}_{\rm IG} (\ell) = \int^{\chi_{\rm h}}_{0} {\rm d}\chi \frac{g^i(\chi)n^j_{\rm s}(\chi) + n_{\rm s}^i(\chi)g^j(\chi) }{\chi^2} P_{\rm GI}(k=\frac{\ell}{\chi}, \chi)\ .
\end{equation}
The $P_{\rm II}$ and $P_{\rm GI}$ are IA power spectra. Throughout the work, we adopt the commonly used nonlinear alignment (NLA) model \citep{Hirata04} to mitigate IA uncertainties,  i.e. assuming the amplitudes of IA power spectra are linearly related to the local density field: 
\begin{equation} 
\label{eq:PII/GI}
\begin{aligned}
P_{\rm II}(k,z) &= A^2(z) P_{\delta}(k,z) \\ 
P_{\rm GI}(k,z) &= A(z) P_{\delta}(k,z)    \\
A(z) &= - A_{\rm IA} C_1 \frac{3 H_0^2 \Omegam}{8 \pi G} D^{-1}(z)\ \left(\frac{1+z}{1+z_0}\right)^{\eta_{\rm IA}} \ .
\end{aligned}
\end{equation}
Here $D(z)$ is the linear growth factor; $C_1$ is the normalization constant being set at $5 \times 10^{-14}\ \Msun^{-1} h^{-2} {\rm Mpc}^3$ \citep{Brown02}; the pivot redshift $z_0$ is being set to 0.62. The nuisance parameters that go into the pipeline for IA marginalization are $A_{\rm IA}$ and $\eta_{\rm IA}$. For a more detailed IA analysis on DES Y1 data see \citet{Samuroff18}.

\subsubsection{Galaxy Clustering}
The location of galaxies traces the underlying matter density field, yet with some unknown bias factor which depends on scales and redshift and on the tracer galaxy population. On large scales, under the simple scale-independent linear bias model, the theoretical prediction for the galaxy-galaxy auto-correlation function in tomographic bin $i$ can be expressed as: 
\begin{equation} 
\label{eq:wp}
\begin{aligned}
w^i(\theta) &= \frac{1}{2 \pi} \int {\rm d}\ell J_0(\ell \theta) C^{ii}_{\delta_{\rm g} \delta_{\rm g} }(\ell)  \  \\
C^{ii}_{\delta_{\rm g} \delta_{\rm g} }(\ell) &=  (b_{\rm g}^i)^2 \int_0^{\chi_{\rm h}} {\rm d}\chi_{\rm l} \frac{(n^i_{\rm l}(\chi_{\rm l}))^2}{\chi_{\rm l}^2} P_{\delta}(k=\frac{\ell}{\chi_{\rm l}},\chi_{\rm l}) \ , 
\end{aligned}
\end{equation}
where $n^i_{\rm l}(\chi_{\rm l})$ is the probability distribution function for the redshift distribution of lens galaxies, and $b_{\rm g}^i$ is the galaxy bias factor for each tomographic bin. 

\subsubsection{Galaxy-Galaxy Lensing}
Galaxy-galaxy lensing, the cross correlation between the position of lens galaxies in bin $i$ and their surrounding matter density field traced by the shear of source galaxies in bin $j$, is modeled as: 
\begin{equation} 
\label{eq:gammat}
\gamma_{\rm t}^{ij}(\theta) = (1+ m^j) \frac{1}{2 \pi} \int {\rm d}\ell J_2(\ell \theta) C^{ij}_{\delta_{\rm g} \gamma}(\ell) \ ,
\end{equation}
where $m^j$ again is the multiplicative shear bias; $J_2$ is the second-order Bessel function. Similarly, the $C^{ij}_{\delta_{\rm g} \gamma}(\ell)$ term has contributions from both pure lensing and IA effects, 
\begin{equation} 
\label{eq:Cgr}
C^{ij}_{\delta_{\rm g} \gamma}(\ell) = C^{ij}_{\delta_{\rm g} \rm G}(\ell) + C^{ij}_{\delta_{\rm g} \rm I}(\ell) \ .
\end{equation}
The lensing term reads
\begin{equation} 
\label{eq:CgG}
C^{ij}_{\delta_{\rm g} \rm G}(\ell) =  b^i_{\rm g} \int^{\chi_{\rm h}}_0 {\rm d}\chi_{\rm l} \frac{n^i_{\rm l}(\chi_{\rm l}) g^j(\chi_{\rm l})}{\chi_{\rm l}^2} P_{\delta}(k=\frac{\ell}{\chi_{\rm l}},\chi_{\rm l}) \ ,
\end{equation}
and the IA term is expressed as 
\begin{equation} 
\label{eq:CgI}
C^{ij}_{\delta_{\rm g} \rm I}(\ell) =  b^i_{\rm g} \int^{\chi_{\rm h}}_0 {\rm d}\chi \frac{n^i_{\rm l}(\chi) n_{\rm s}^j(\chi)}{\chi^2} P_{\rm GI}(k=\frac{\ell}{\chi},\chi) \ , 
\end{equation}
with the IA power spectrum $P_{\rm GI}$ being defined in Eq.~\eqref{eq:PII/GI}.

\bigbreak

\noindent Finally, throughout this work, the uncertainty in the photometric redshifts is modeled as a constant shift of the initial redshift probability distribution function $n^i_{\rm pz}(z)$, for both source and lens galaxies, in each tomographic bin. 
\begin{equation}
\label{eq:photoz}
n^i_{\rm s}(z) = n^i_{\rm s, pz}\ (z - \Delta z_{\rm s}^i)\ \ \ ; \ \ \ n^i_{\rm l}(z) = n^i_{\rm l, pz}\ (z - \Delta z_{\rm l}^i)
\end{equation}

\begin{table}
\caption{Parameters and priors used to run the likelihood analyses. Flat($a$, $b$) denotes a flat prior in the range given while Gauss($\mu,\sigma$) is a Gaussian prior with mean $\mu$ and width $\sigma$. The third column summarizes the fiducial parameter values we used to generate mock data vectors and to construct PCs. The fiducial values are chosen to be consistent with the posterior constraints from the fiducial \LCDM model of DES Y1 3$\times$2pt analyses \citep{DES18}. The fiducial photo-z and shear calibration parameters are set at the peak of the Gaussian prior for the purpose of running likelihood simulations.  }
\begin{center}
\begin{tabular}{clc}
\hline
Parameter  		&     Prior     		& 	Fiducial Value	\\  \hline
\multicolumn{3}{c}{{\bf Cosmology}} \\
$\Omegam$		& Flat (0.1, 0.9) 		& 	0.3	\\
$A_{\rm s}$		& Flat ($5\times10^{-10}$, $5\times10^{-9}$) 		&	$2.19\times10^{-9}$	\\
$n_{\rm s}$		& Flat (0.87, 1.07) 	&	0.97	\\
$\Omegab$		& Flat (0.03, 0.07) 	&	0.048	\\
\multirow{2}{*}{$\Omega_{\nu} h^2$} & baseline\ :\ Flat ($5\times10^{-4}$, $0.0013$) 	&	\multirow{2}{*}{0.00083}	\\
			        				& Y1 fiducial\ :\ Flat ($5\times10^{-4}$, $0.01$) 		&	   		 \\
$h$				& Flat (0.55, 0.91)  	&	0.69  	\\
\hline
\multicolumn{3}{c}{{\bf Lens Galaxy Bias}} \\
$b_{\rm g}^1$   	& Flat (0.8, 3.0)		& 1.53 	\\
$b_{\rm g}^2$   	& Flat (0.8, 3.0)		& 1.71 	\\
$b_{\rm g}^3$   	& Flat (0.8, 3.0)		& 1.70 	\\
$b_{\rm g}^4$   	& Flat (0.8, 3.0)		& 2.05 	\\
$b_{\rm g}^5$   	& Flat (0.8, 3.0)		& 2.14 	\\
\hline
\multicolumn{3}{c}{{\bf Lens \photoz\ shift}}  \\
$\Delta z^1_{\rm l}$  & Gauss ($0.008, 0.007$) 		& 0.008	\\
$\Delta z^2_{\rm l}$  & Gauss ($-0.005, 0.007$)	& -0.005	\\
$\Delta z^3_{\rm l}$  & Gauss ($0.006, 0.006$) 		& 0.006	\\
$\Delta z^4_{\rm l}$  & Gauss ($0.0, 0.01$) 		& 0.0		\\
$\Delta z^5_{\rm l}$  & Gauss ($0.0, 0.01$) 		& 0.0 	\\
\hline
\multicolumn{3}{c}{{\bf Source \photoz\ shift}} \\
$\Delta z^1_{\rm s}$  & Gauss ($-0.001, 0.016$) 	&  -0.001  \\
$\Delta z^2_{\rm s}$  & Gauss ($-0.019, 0.013$) 	&  -0.019   \\
$\Delta z^3_{\rm s}$  & Gauss ($+0.009, 0.011$) 	&  0.009  \\
$\Delta z^4_{\rm s}$  & Gauss ($-0.018, 0.022$) 	&  -0.018 \\
\hline
\multicolumn{3}{c}{   {\bf Shear calibration} (\bf \textsc{metacalibration})  }  \\
$m^1$ & Gauss ($0.012, 0.023$)	& 0.012 \\
$m^2$ & Gauss ($0.012, 0.023$)	& 0.012 \\
$m^3$ & Gauss ($0.012, 0.023$)	& 0.012 \\
$m^4$ & Gauss ($0.012, 0.023$)	& 0.012 \\
\hline
\multicolumn{3}{c}{{\bf Intrinsic Alignment} } \\
$\AIA$   	& Flat ($-5,5$) &	 0.45	\\
$\etaIA$   	& Flat ($-5,5$) &	 -1.0	\\
\hline
\multicolumn{3}{c}{{\bf Baryon PC amplitude}} \\
\multirow{2}{*}{$Q_1$}  &  baseline\ :\ Flat ($-3,12$)  & \\
				    & 	informative\ :\ Flat ($\ 0,\ 4$) & \\
$Q_2$			    & Flat ($-2.5,2.5$) & \\						       
\label{tb:params}
\end{tabular}
\end{center}
\end{table}

\subsection{PC Decomposition to model baryonic effects}
\label{subsec:PCA_overview}
    
We adopt the principal component (PC) decomposition technique to model baryonic effects for small-scale cosmic shear \citep{Eifler15}. The basic idea of this technique is to perform principal component analysis (PCA) on the difference of the theoretical model vectors (the $3 \times 2$pt vectors for this work) between hydrodynamical and DMO simulations, for several baryonic scenarios.
To construct the baryon-contimanted cosmic shear correlation functions, we use Eq.~\eqref{eq:Pk_ratio} to derive the underlying baryonic power spectra that go into integration (see \S\ref{subsec:mock_data}). 

The resulting dominant PC modes then serve as a flexible basis set to account for possible baryonic effects in both spatial and redshift dimensions via the angular bins and tomographic information. In \citetalias{Huang19}, we validate this method assuming an LSST-like cosmic shear experiment. 
We further improve the efficiency of this method by imposing a covariance-driven weighting factor when performing PCA, which is referred to as method C in \citetalias{Huang19}. Below we briefly summarize the formalism of this method.

Let $\M$ be a DMO-based theoretical $3\times2$pt model vector, and $\B_{x}$ be a model vector contaminated with baryonic scenario $x$, computed by replacing the matter power spectrum via Eq.~\eqref{eq:Pk_ratio} (see \S\ref{subsec:mock_data} for detail). We first build a difference matrix $\bm \Delta$
\begin{equation} \label{eq:DiffMatrix}
  \begin{aligned}
  {\bm \Delta} = \left[
    \begin{array}{cccc}
      \B_{1}-\M    & \B_{2}-\M    & \ldots & \B_{N_{\rm sim}}-\M    \\
    \end{array}
  \right]_{N_{\rm data} \times N_{\rm sim}} \ .
   \end{aligned}
  \end{equation}
  Each column of $\bm \Delta$ is a difference vector, $\B_x-\M$, with 630 elements (\S\ref{subsec:DES_data}), computed with the cosmology and the nuisance parameters being set to the fiducial values listed in Table~\ref{tb:params}.

Next we use the Cholesky decomposition on the data vector covariance matrix, to find the square root of the covariance
  \begin{equation} \label{eq:chy_de}
    \C = \L \L^{\t}\ .
    \end{equation}

  We use $\invL$ to build a noise-weighted difference matrix $\bm \Delta_{\rm ch}$, and apply singular value decomposition (SVD) to $\bm \Delta_{\rm ch}$
  \begin{equation} \label{eq:DiffMatrix_chy}
    \begin{aligned}
    {\bm \Delta_{\rm ch}} &= \invL {\bm \Delta} \\
    &= \invL 
    \left[
      \begin{array}{cccc}
        \B_{1}-\M    & \B_{2}-\M    & \ldots & \B_{N_{\rm sim}}-\M    \\
      \end{array}
    \right]_{N_{\rm data} \times N_{\rm sim}} \\
     &  = \U_{\rm ch} \ \bm \Sigma_{\rm ch}\ \V_{\rm ch}^{\t} \ ,
     \end{aligned}
    \end{equation}
    where $\U_{\rm ch}$ and $\V_{\rm ch}$ are square unitary matrices with dimensions of $N_{\rm data}\times N_{\rm data}$ and $N_{\rm sim}\times N_{\rm sim}$ respectively. $\bm \Sigma_{\rm ch}$ is a diagonal matrix with the singular values populating the diagonal in descending order.  

The first $N_{\rm sim}$ columns of the $\U_{\rm ch}$ matrix form a set of PC bases, ${\mathbf v}_{{\rm PC}, i}$, that can be used to fully span the baryonic features of our training simulations. For a given baryonic scenario $x$, we have
\begin{equation} \label{eq:invLB-M}
\invL (\bm B_{x} - \M) = \sum_{i = 1}^{N_{\rm sim}} Q_i\ {\mathbf v}_{{\rm PC}, i} \ .
\end{equation}

With the derived PCs, we can generate a baryonic model that utilizes PC amplitudes $Q_i$ to simulate possible baryonic behaviors. 
\begin{equation} \label{eq:Mbary}
\M_{\rm bary}(\pco, \pnu, \Q ) = \M(\pco, \pnu) + \sum_{i = 1}^{n} Q_i\ \L \cdot {\mathbf v}_{{\rm PC}, i} \ .
\end{equation}
Here $n$ specifies the number of PC amplitudes/PC modes used to model the baryonic effect, and $n \leq N_{\rm sim}$. The operation of $\L \cdot {\mathbf v}_{{\rm PC}, i}$ transforms the PC mode back to the same basis as $\M$.

Note that although we pass the full 3$\times$2pt vector in Eq.~\eqref{eq:DiffMatrix_chy} to perform PCA, the deviations from the DMO scenario are extremely small for the galaxy-galaxy lensing and galaxy clustering parts because of their conservative scale cuts. Therefore, the PCs mostly account for baryonic effects in small-scale data points of cosmic shear (see Fig.~\ref{fig:Mpar_1sigma_sym} for the fractional change of model vector when varying $Q_1$).

\subsubsection{Input hydrodynamical scenarios for PC construction}
\label{subsec:PC_construction}

We will use the Illustris and the Eagle scenarios as the conservative and optimistic validation scenarios for our PCA baryon mitigation model.

As mentioned before, we exclude the considered scenario in the baryon PC basis set, hence we are building two PC bases for this exercise, one excluding Eagle and the second one excluding Illustris.

The first PC set is constructed with 10 hydrodynamical scenarios: MB2, Horizon-AGN, TNG100, Eagle, three variants of cosmo-OWLS, and three variants of BAHAMAS scenarios with different AGN feedback strength. We will use this basis set to mitigate baryonic effects for our Illustris and DMO mock data vectors (see \S\ref{subsec:mock_data}), and for the real DES Y1 observational data vector.

The second PC basis set is constructed with the same scenarios as the first, with the Eagle scenario being excluded. When performing our analyses on the Eagle mock data vector, we will use the second PC set as bases to conduct baryon mitigation. The reasoning for this design can be understood in  Eq.~\eqref{eq:invLB-M}. If using the first PC set to perform marginalization on Eagle, the first PC set is guaranteed to be able to describe Eagle by construction.

\begin{figure}
\begin{center}
\includegraphics[width=0.47\textwidth]{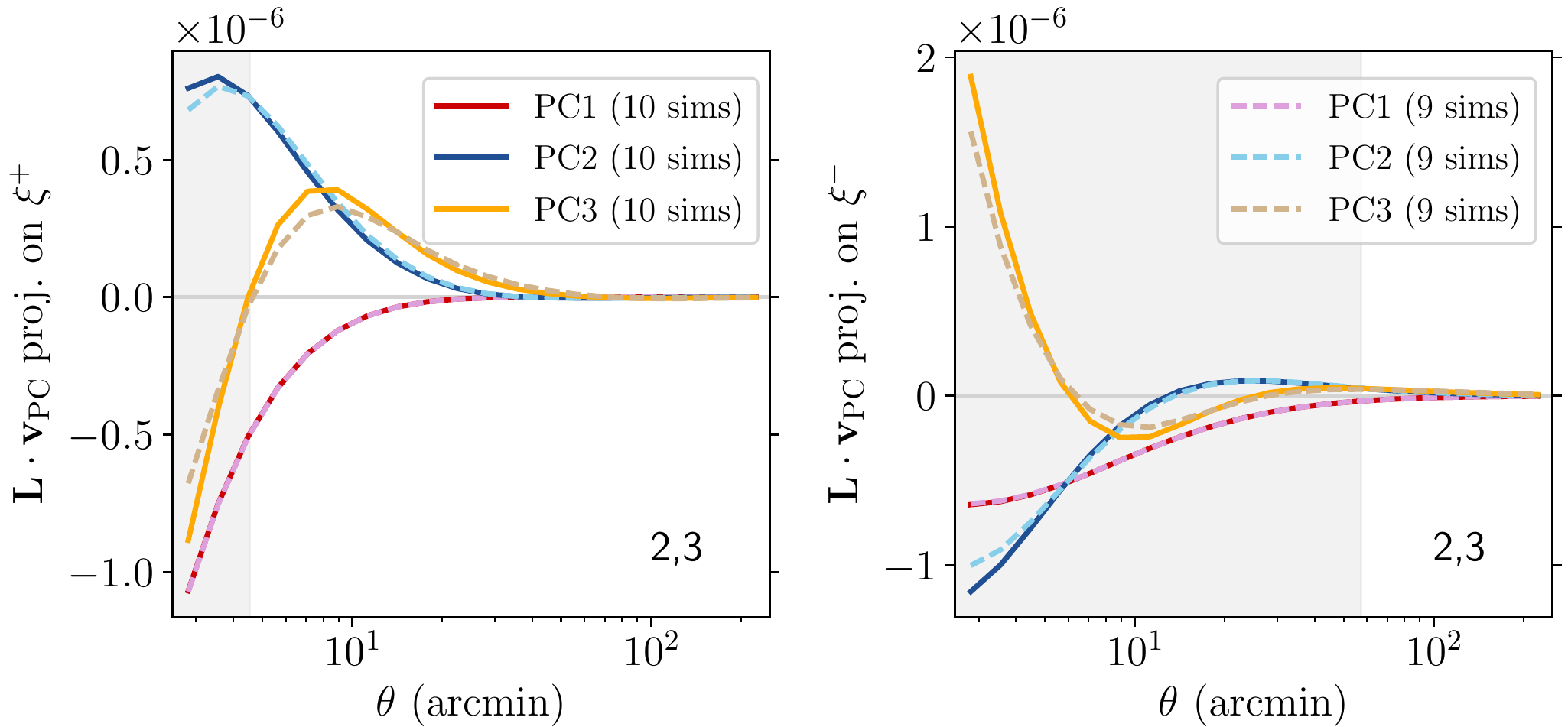}
\caption{The principal components used in our baryon model (Eq.~\ref{eq:Mbary}).
Here we show the first three $\L \cdot {\mathbf v}_{{\rm PC}}$ components projected on the cosmic shear correlation functions in the cross tomographic bin (2,3). The solid curves indicate PCs constructed based on 10 hydrodynamical scenarios, which are used when analyzing the DES data, and validating our pipeline on mock data construted from the Illustris and the DMO scenarios. 
The dashed curves are constructed from 9 hydrodynamical scenarios, which are used when validating on the Eagle mock data (see \S{\ref{subsec:PC_construction}} for detail). The gray shaded backgroud regions highlight the angular scales that excluded in the original Y1 cosmic shear analysis. \textit{In this work, we include these small-scale cosmic shear data points, and use the PCs as flexible bases to span uncertainties of baryons in cosmic shear.}}
\label{fig:PCs}
\end{center}
\end{figure}

Figure~\ref{fig:PCs} provides a visualization of $\L \cdot {\mathbf v}_{{\rm PC},i}$ in projection on the $\xi_{\pm}$ observables in the cross tomographic bin (2,3), for our two sets of PC bases. As shown, these two sets of PC modes turn out to be quite similar.

\subsection{Likelihood Analysis}
\label{subsec:likelihood_analysis}

We infer the posterior probability distribution of cosmological ($\pco$) and nuisance parameters ($\pnu$) via Bayes' theorem:
\begin{equation}
\label{eq:Bayes}
P(\pco, \pnu | \D) \propto L(\D | \pco, \pnu) P_{\rm prior} (\pco, \pnu)\ ,
\end{equation}
with the prior probability distribution for each of the parameter defined in Table~\ref{tb:params}. 

\begin{figure}
\begin{center}
\includegraphics[width=0.47\textwidth]{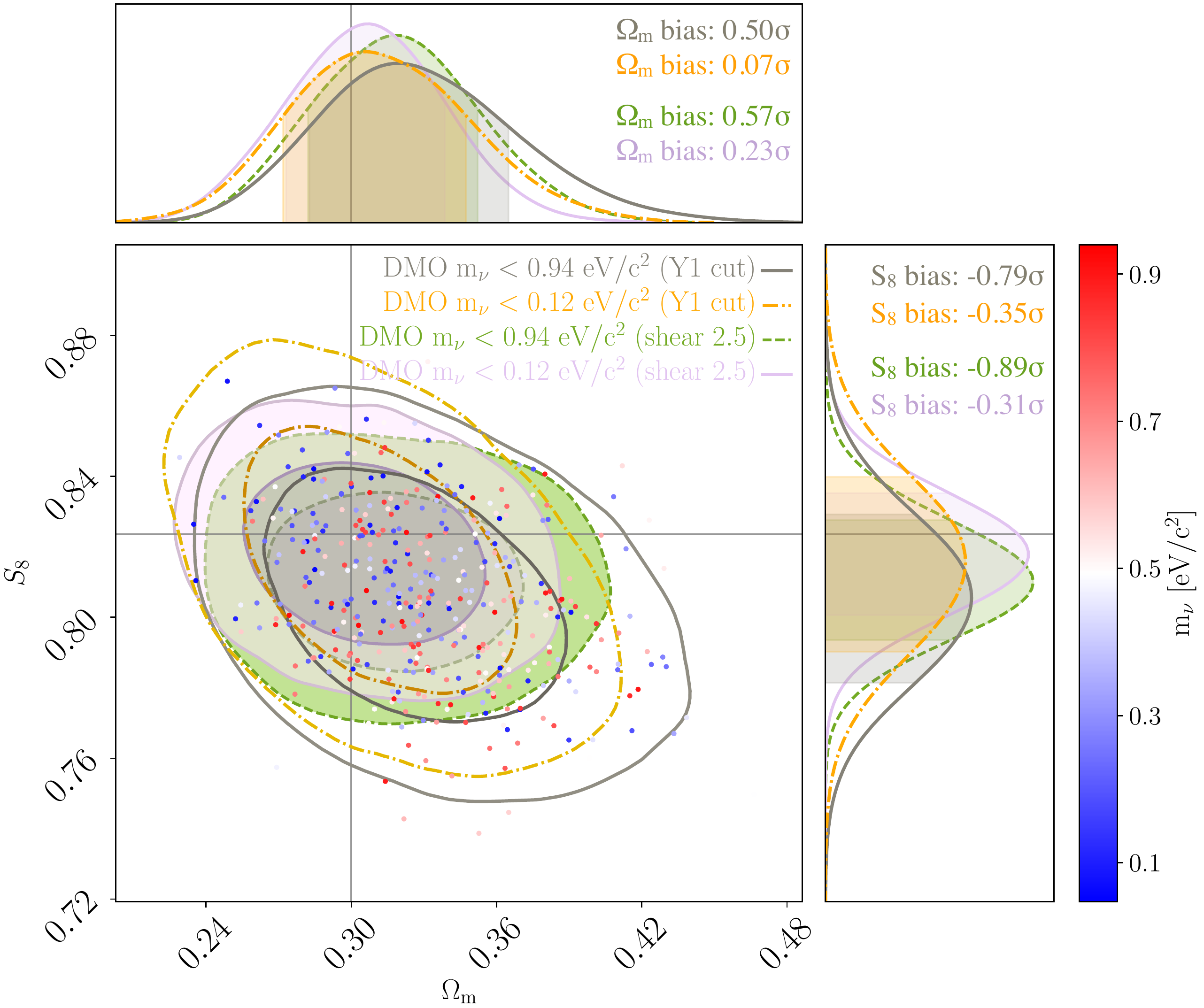}
\caption{Cosmological parameter constraints from simulated likelihood analyses subjected to different choices of neutrino mass priors. All chains shown here are based on the DMO mock 3$\times$2pt analyses. The gray (orange) contours indicate the case with fiducial Y1 scale cuts, and with wide (narrow) neutrino prior applied. The shaded green (pink) contours are the case when extending cosmic shear to 2.5$\arcmin$ in the 3$\times$2pt mock data (without performing baryon marginalization), subjected to the Y1 fiducial (narrow) neutrino prior.
Here, and in all 2D posterior plots below, the contours depict the 68\% and 95\% confidence levels.
\textit{The parameter biases of $\Omegam$ and $S_8$ decrease when narrowing the neutrino prior.} 
The colored dots are randomly selected samples in the wide neutrino prior chain (orange curves) with the neutrino mass colored as indicated in the sidebar. 
 \textit{Higher neutrino mass tends to suppress the clustering amplitude of matter. The posterior of $\Omegam$ is thus biased high to compensate for that.}
 } 
\label{fig:neutrino_prior} 
\end{center}
\end{figure}

\begin{figure}
\begin{center}
\includegraphics[width=0.47\textwidth]{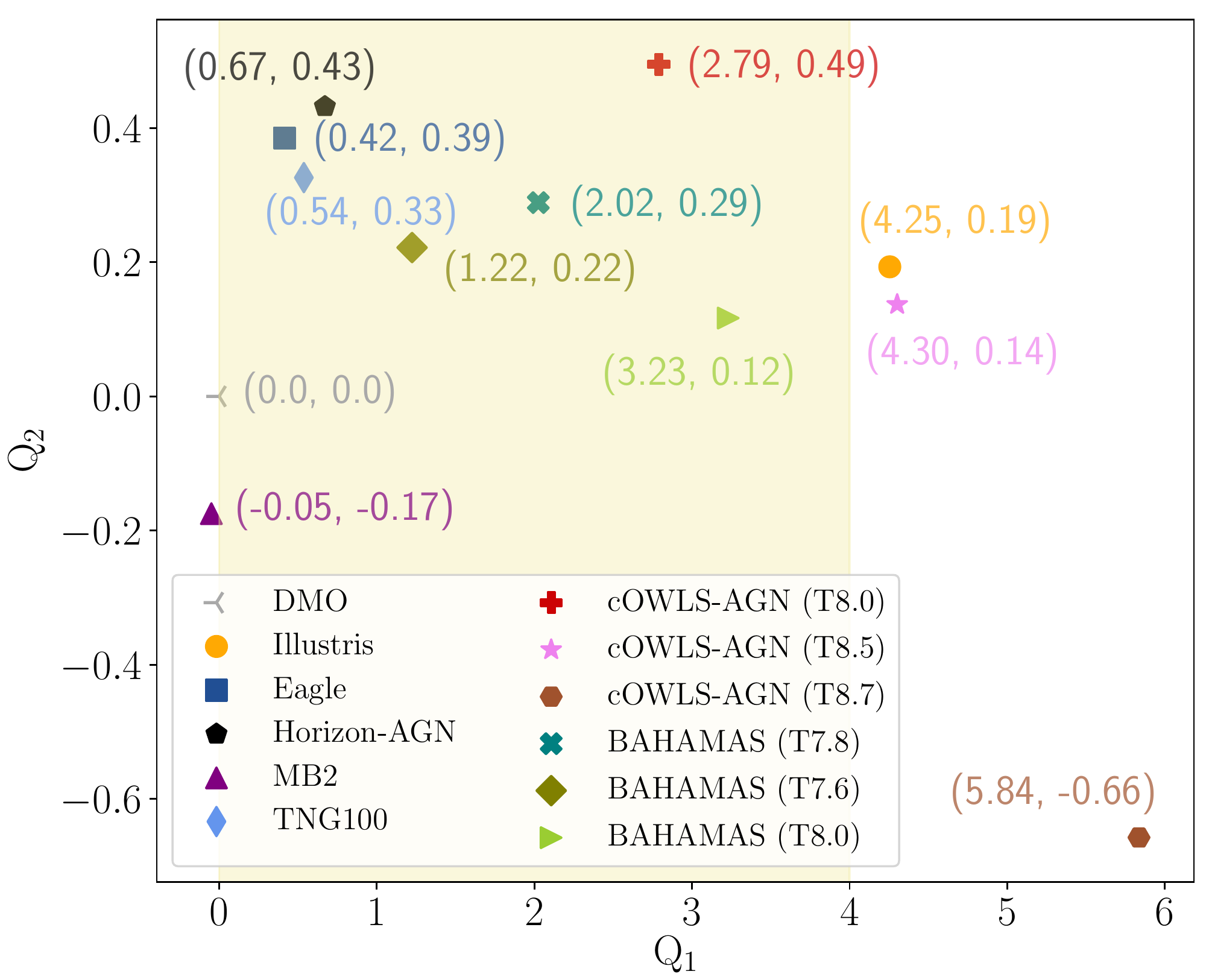}
\caption{The expected PC amplitudes $Q_1$, $Q_2$ for baryonic scenarios considered in this work, with PCs being constructed using 10 hydrodynamical simulations as detailed in \S\ref{subsec:PC_construction}. Our choice of priors for $Q_{1,2}$ are based on the range of values from these baryonic scenarios. The yellow band highlights the range of informative prior on $Q_1$, Flat(0,4), which is bounded by the Illustris scenario. 
\textit{As shown in Fig.~\ref{fig:PCs} on the features of PC modes, the larger $Q_1$ value indicates stronger suppression of matter clustering at small scales. The $Q_2$ parameter further provides a higher order correction.}
}
\label{fig:Q1Q2_Scatter}
\end{center}
\end{figure}

The priors for our baseline analysis are chosen to be mostly the same as DES Y1 \citep{DES18}, with an exception of the upper limit of the neutrino mass prior. We now discuss the priors in more detail.

\subsubsection{Prior for neutrino mass} 
\label{subsec:neutrino_prior}

Instead of applying a non-informative upper limit on the sum of neutrino masses ${\Sigma m_{\nu}} < 0.94$ eV/$c^2$ (i.e. $\Omega_\nu h^2 < 0.01$) as in DES Y1, we adopt an upper limit of $\Sigma {m_{\nu}} < 0.12$ eV/$c^2$ ($\Omega_\nu h^2 < 0.0013$) as our baseline analysis. This upper limit is based on the latest 95\% constraint from $Planck$ TT, TE, EE+lowE+lensing+BAO \citep{Planck18} and has the advantage that it reduces biases in the 1D projected posterior probabilities of the relevant cosmological parameters. 

As shown in Fig.~\ref{fig:neutrino_prior}, for the simulated likelihood analysis with DES Y1 scale cuts, the wide Y1 neutrino prior leads to a $\sim$0.8$\sigma$-level bias in $S_8$ (gray contour); while the case with an informative neutrino prior only has a $0.35\sigma$ bias in $S_8$. The bias is caused by the asymmetric coverage of the neutrino prior around the fiducial neutrino value, as discussed in \citet{Krause17}. 
Since DES Y1 data do not have significant constraining power on neutrino masses, marginalizing over neutrino mass preferentially allows many scenarios with increased neutrino mass, which leads to a net suppression in structure growth. The $\Omegam$ posterior is then biased high to compensate for that.

This neutrino prior-induced bias becomes more significant when including small-scale data in the analysis, due to the smaller uncertainties  (gray versus green contours in Fig.~\ref{fig:neutrino_prior}) and due to the fact that small-scale data are more sensitive to the neutrino mass. We thus place a narrower limit on the neutrino mass prior as our baseline analysis, and the $S_8$ bias is reduced to $\sim 0.3\sigma$ (shaded pink contours). 
For the purpose of comparison, we will also present and discuss the result with the original Y1 wide neutrino prior.

\subsubsection{Priors for baryonic parameters} 
\label{subsec:baryon_prior}

The theoretical PC amplitude $Q_{i}$ for each hydrodynamical scenario $x$ can be computed by taking the inner product of the weighted difference vector, $\invL (\bm B_{x} - \M)$, with the PC mode PC$_i$ (see Eq.~\eqref{eq:invLB-M}). Fig.~\ref{fig:Q1Q2_Scatter} presents the expected $Q_{1,2}$ values for all hydrodynamical scenarios considered in this work.

An increase in $Q_1$ mostly controls the amount of suppression on small scales, whereas higher order PC amplitudes $Q_{i \geq 2}$ provide corrections on baryonic effects that can also impact larger scales (see Fig.~\ref{fig:PCs}).

We adopt two choices of priors for the baryonic parameters.
\begin{itemize}
  \item baseline\ \ \ \ \ \ : $Q_{1} \in$ Flat(-3, 12)\ \ \ ; \ \ $Q_{2} \in$ Flat(-2.5, 2.5)
  \item informative\ : $Q_{1} \in$ Flat(\ 0,\ \ \ 4) \ \ ; \ \ $Q_{2} \in$ Flat(-2.5, 2.5)
\end{itemize}

The baseline priors are extremely conservative and allow for the data to entirely self-calibrate the baryonic effects. Looking at Fig.~\ref{fig:Q1Q2_Scatter} we see that they are significantly larger than the spread of $Q_{1,2}$ for all hydrodynamical scenarios. 

The informative prior of $Q_{1}$ is highlighted in the yellow band of Fig.~\ref{fig:Q1Q2_Scatter}, and we consider this prior range to be well-motivated if one considers including a minimal amount of external information from the simulation literatures in our analysis. Specifically, \cite{Haider16} found that the radio-mode AGN feedback in Illustris is too strong such that too much gas is heated and ejected, leading to insufficient baryons in galaxy groups compared with observations. Also in \cite{LeBrun14}, the cosmo-OWLS T8.5 and T8.7 scenarios are a poor match to several X-ray observables. It is thus reasonable to use the Illustris scenario as an upper bound on the level of feedback strength that our Universe could possibly reach and to adopt a corresponding prior in our analysis. As we will see later, this informative, but well motivated prior, will increase the amount of information we gain on cosmology by adding small-scale cosmic shear data in DES Y1.

\bigbreak

For our likelihood analyses, we adopt a Gaussian likelihood: 
\begin{equation} 
\label{eq:likelihood}
L(\D | \pco, \pnu) \propto \exp \biggl( -\frac{1}{2} \underbrace{\left[ (\D -\M)^t \, \C^{-1} \, (\D-\M) \right]}_{\chi^2(\pco, \pnu)}  \biggr) \ .
\end{equation}
As discussed in \citet{Lin19}, the impact of non-Gaussianity in the likelihood is estimated to be negligible in current and future cosmic shear surveys.

The 3$\times$2pt covariance matrix $\C$ is computed using the \textsc{CosmoLike} package \citep{Krause17b}, which calculates the relevant four-point functions in the halo model. The analytic form of the covariance matrix and relevant validation for DES Y1 is detailed in \citet{Krause17}, with updates provided in \citet{Troxel18b} to address the effect of survey geometry and the uncertainty in the  multiplicative shear bias calibration. 
For simplicity, we do not consider potential baryonic effects when computing the covariance matrix. As discussed in previous works, neglecting baryonic effects in the covariance matrix has little impact on the cosmological inference for stage IV weak lensing surveys \citet{Schneider20, Barreira19}, and should be negligible for DES.

We use the \texttt{emcee} package \citep{Foreman-Mackey13}, which relies on the affine-invariant ensemble sampling algorithm \citep{goodman2010ensemble}, to sample the parameter space. We run MCMC (Markov Chain Monte Carlo) chains to 2.5 million steps, and then discard the first 1.25 million steps as burn-in. We have visually checked the convergence of MCMC chains by ensuring that the 1D and 2D posterior distributions for all parameters  are consistent with the results of a chain with 5 million steps out to $3\sigma$ confidence intervals.

\subsection{Blinding Strategy}
\label{subsec:blinding}
Our blinding strategy aims to shield against ``confirmation bias", i.e. stopping the search for new systematics or better parameterizations of existing systematics when the result matches the expectation. 
There are differences between our analysis and the DES Y1 analysis choices described in \citet{Krause17}, and these analysis differences will drive those in the respective blinding strategies.
In particular we include small scales in cosmic shear (down to 2.5$\arcmin$), we add a corresponding parameterization for baryonic physics uncertainties, and we use a different prior for the neutrino mass parameter. 
Beyond these differences, we follow the \citet{Krause17} choices; in particular, we do not reassess scale cuts for galaxy-galaxy lensing and galaxy clustering, or other model parameterizations and priors. This is justified given that our constraining power is very similar to that of \citet{DES18}, and even when we use informative priors on baryonic physics we expect a 20\% information increase at most (see \S\ref{subsec:informative_baryon_prior}).
Based on these considerations, our blinding strategy proceeds as follows:
\begin{enumerate}
    \item We develop our pipeline completely independently of the data vector, i.e.\ we run 100+ simulated likelihood analyses to stress-test our pipeline. We use different data vectors, different prior settings, and different modeling settings until we converge to the setup described in Table~\ref{tb:params}. We describe this process in \S\ref{subsec:likelihood_analysis} and results in \S\ref{sec:pipeline}.
    \item Our pipeline is a modified version of the DES Y1 \textsc{CosmoLike} pipeline; we performed a comparison with the latest \textsc{CosmoLike} version (which has undergone testing and validation for DES Y3) and have reached an excellent agreement at the level of $\Delta \chi^2=0.0005$ and $\Delta \chi^2=0.0006$ for model vectors with the original scale cuts used in \citet{DES18} and with the new scale cuts used in this paper, respectively. The residual uncertainties are due to small modifications in the interpolation routines that were incorporated between Y1 and Y3. The version used in this work is tagged as 'Huang2020' in the 'cosmolike$\_$core' github repository of the \textsc{CosmoLike} github organization.  
    
    \item We described all our pipeline tests and the code comparison to an internal review panel within the DES collaboration and only replaced the simulated data vector with the actual data after their sign-off. 
\end{enumerate}

The data constraining results presented in \S \ref{sec:baryon_constraint} and \S \ref{sec:cosmology_constraint} are unaltered post unblinding.

\section{Likelihood Simulation Results}
\label{sec:pipeline}

In this section we present our simulated likelihood analysis results for the three mock data vectors of baryonic scenarios, DMO, Eagle, and Illustris, in order to design and understand the expected performances of our baryon mitigation pipeline. 
DMO is the best-case scenario for which we know in advance that the resulting cosmological inference should not be biased, regardless of whether baryon mitigation is performed. With its strong feedback, the Illustris simulation serves as a conservative scenario in our pipeline validation; such strong feedback is largely ruled out by observations already \citep{Haider16}. 
The Eagle scenario has significantly weaker feedback, so its deviation from DMO is relatively small and it serves as an optimistic scenario in our pipeline validation.

\begin{figure*}
  \begin{center}
  \includegraphics[width=1\textwidth]{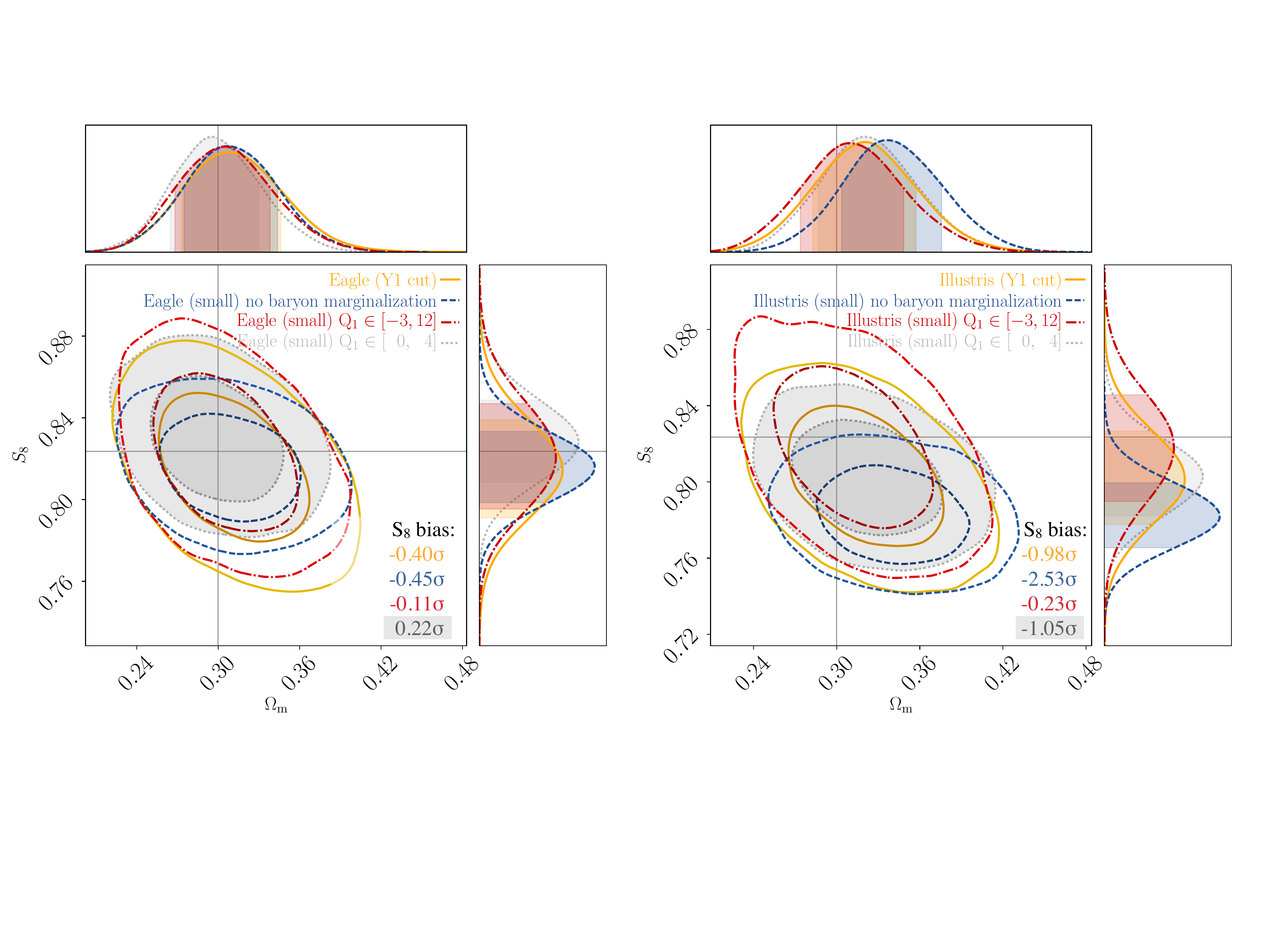}
  \caption{
Results with the analysis setup where we fit the 3x2pt observables while marginalizing over baryonic physics on two hydrodynamical simulations. The two panels show posterior constraints on $\Omegam$ and $S_8$ for the Eagle (left) and the Illustris (right) mock data.
  The yellow contours indicate the result when the DES Y1 scale cut is applied. The blue (red) contours show the result when extending cosmic shear to 2.5$\arcmin$, but without (with) marginalization on baryonic parameter $Q_1$. The analysis result when adopting the informative $Q_1$ prior is indicated in gray shaded contours. The marginalized 1D constraint in $S_8$ with $1\sigma$ error is spelled out in the lower right corners of the plots.} 
  \label{fig:cosmology_hydrosims}
 \end{center}
\end{figure*}

As an overview, in Fig.~\ref{fig:cosmology_hydrosims}, we show the posterior distributions of $\Omegam$ and $S_8$ with the input mock 3$\times2$pt data from the Eagle (left panel) and  Illustris (right panel) scenarios. We compare the fiducial DMO case (filled grey contours) with: 
\begin{enumerate}
	\item applying DES Y1 scale cuts (yellow contours), 
	\item extending cosmic shear to 2.5$\arcmin$ but without introducing an extra parameter to marginalize over baryonic physics (blue contours), 
	\item same as (ii), but marginalizing over $Q_1$ with our baseline prior Flat(-3, 12) (red contours), 
	\item same as (iii), but applying an informative prior  Flat(0, 4) on $Q_1$ (gray shaded contours).
\end{enumerate}
Below we will investigate the posterior distributions  on these simulated likelihood analyses (shown in Fig.~\ref{fig:cosmology_hydrosims}), to understand the potential outcomes when applying our pipeline on real data.

\subsection{Number of PC modes to be marginalized over given DES Y1 constraining power}
\label{subsec:NPCmodes_determination}

To determine how many PC modes are needed in Eq.~\eqref{eq:Mbary} to account for baryons when pushing cosmic shear to 2.5$\arcmin$ given DES Y1 statistical power,
we increase the available degrees of freedom by increasing the number of PC amplitudes $Q_i$ when running likelihood simulations and track the resulting posterior distributions. 

\subsubsection{The residual bias after marginalization}
\label{subsec:residual_bias}

\begin{figure}
\begin{center}
\includegraphics[width=0.47\textwidth]{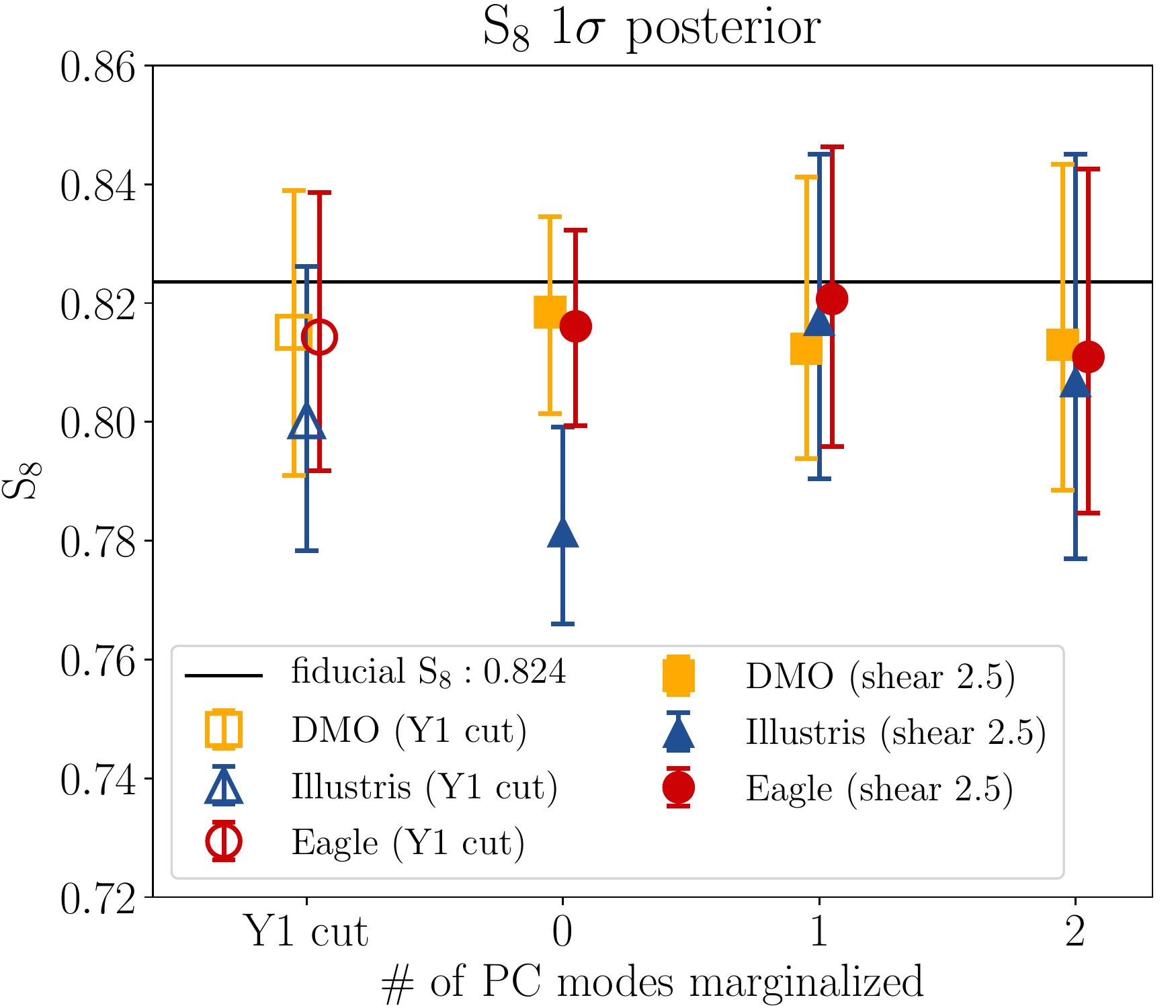}
\caption{The marginalized 1D $S_8$ posterior constraints for our baseline likelihood simulations with baryonic scenarios of DMO, Eagle and Illustris. Each marker's center, lower, and upper error bars indicate the median, the 16th and the 84th percentiles of marginalized 1D posteriors. The open markers are results of 3$\times$2pt mock data vectors subjected to the original DES Y1 scale cuts, while the filled markers are results when extending the cosmic shear data points to 2.5$\arcmin$ (\S\ref{subsec:data}), with different choices for the number of marginalized PC amplitudes $Q_i$ to account for baryonic effects. 
\textit{Marginalizing over 1 PC mode is sufficient to account for baryonic effects to within $\sim 0.2\sigma$ under the statistical power of DES Y1, for all baryonic scenarios considered here.}
}
\label{fig:S8_1D}
\end{center}
\end{figure}

Figure~\ref{fig:S8_1D} summarizes the marginalized 1D $S_8$ posterior constraints for our likelihood simulations (as shown in Fig~\ref{fig:cosmology_hydrosims} for the case of the Eagle and Illustris scenarios). 

We use the DMO results as a baseline for understanding the level of parameter projection effects, i.e., the parameter biases as revealed in the marginalized posterior constraints. Parameter degeneracies in the high-dimensional space may lead the 1D and 2D projected posteriors to peak at biased positions, and for parameters where the data are not sufficiently constraining the posterior can peak at biased values due to prior volume effects (e.g., the neutrino prior issue discussed in \S\ref{subsec:neutrino_prior}). As indicated by the yellow square markers, we observe that the projection effects would cause $\approx$ $0.3\sim0.5\sigma$ biases in the $S_8$ constraints under our baseline setting (see Table~\ref{tb:params}), which we should keep in mind when interpreting tensions between different experiments using distances in projected parameter spaces.

When performing analyses with the Y1 scale cut (open markers) without marginalizing over baryonic physics, we find a residual $\sim0.9\sigma$ bias in $S_8$ for the Illustris scenario. This is because the Y1 cosmic shear scale cut is determined based on the cOWLS-AGN $\Delta T_{\rm heat} = 10^8$ scenario, which is less intense compared with the feedback effect of Illustris (see Fig.~\ref{fig:Pk_Ratio}).
When including small-scale cosmic shear data points in the 3$\times$2pt analyses (filled markers), for weaker baryonic scenarios like Eagle, we find that even without marginalization the $S_8$ bias can still be within $0.5\sigma$. Using a strong feedback scenario like Illustris as the most pessimistic limit, we conclude that marginalizing over a single PC mode would be sufficient to account for baryonic effects to within $\sim0.2\sigma$, which is well within the referential bias level set from the DMO case.

\subsubsection{The degradation on parameter constraints after marginalization}

\begin{figure}
\begin{center}
\includegraphics[width=0.47\textwidth]{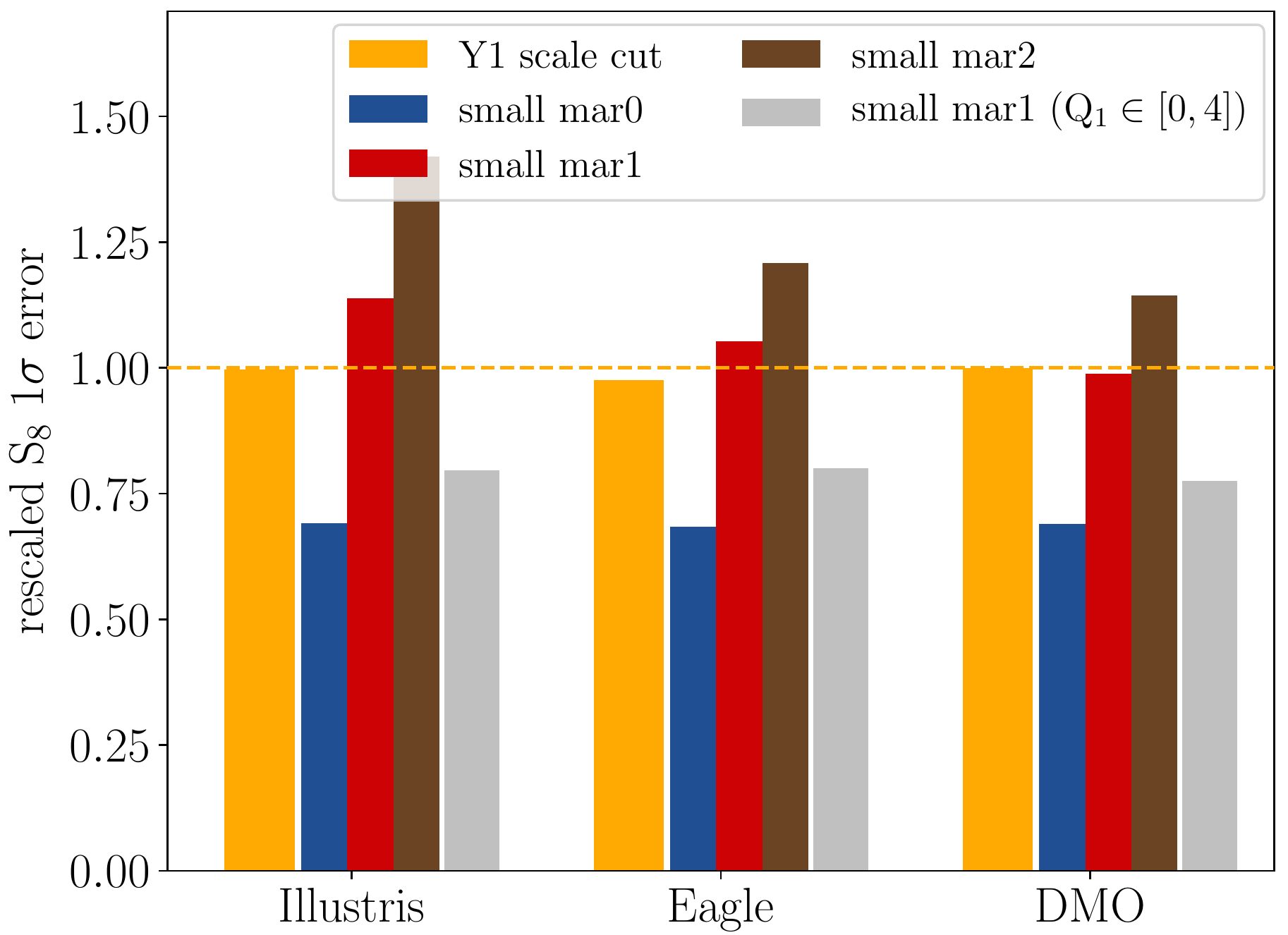}
\caption{The rescaled $S_8$ 1$\sigma$ error in the simulated likelihood analyses: $\sigma/\sigma_{\rm Y1\ cut\ DMO}$. 
The horizontal yellow dashed line indicates the condition when $\sigma=\sigma_{\rm Y1\ cut \ DMO}$.
Starting from left to right, the first/second/third groups indicate results when using the Illustris/Eagle/DMO scenarios as mock data in our likelihood simulations. The yellow bars are for $S_8$ errors derived with Y1 scale cuts applied. The blue/red/brown bars are the results with the cosmic shear data points extended to 2.5$\arcmin$, and with 0/1/2 PC mode(s) being marginalized. The gray bars indicate the results when marginalizing over $Q_1$ with an informative prior range. 
\textit{For the baseline analysis setting with small-scale cosmic shear included, marginalizing over 1 PC mode leads to similar constraining power in $S_8$ compared with the result with Y1 scale cuts being applied. When adopting an informative prior on $Q_1$, a $\sim 20\%$ improvement in $S_8$ is expected.}
}
\label{fig:S8_error}
\end{center}
\end{figure}

Small-scale cosmic shear data points provide additional cosmological information, but some of the information will be lost after accounting for uncertainties in baryonic physics. Here, we explore the expected degradation on parameter constraints in DES Y1 within the PCA framework, subject to our choices on the number of marginalization parameters for baryons. 

Figure~\ref{fig:S8_error} shows the rescaled $S_8$ 1$\sigma$ error for our likelihood analyses (as shown in Fig.~\ref{fig:cosmology_hydrosims} for the cases of the Eagle and Illustris scenarios). Starting from left to right, the first/second/third group is for the Illustris/Eagle/DMO mock data vectors when running likelihood simulation. The yellow bars are for $S_8$ errors derived with Y1 scale cuts applied. The blue/red/brown bars are the results with the cosmic shear data points extended to 2.5$\arcmin$, and with 0/1/2 PC mode(s) being marginalized.
 This figure confirms our expectation that after marginalizing over one PC mode (red bars), the resulting $S_8$ constraint should be similar to the result with conservative scale cuts being applied (yellow bars). Marginalizing over two PC modes (brown bars) should lead to 20\%$\sim$30\% larger errors in $S_8$, depending on the baryonic scenarios. 

In conclusion, we do not expect to gain extra cosmological information from small-scale cosmic shear data points when using the our wide baseline prior to account for baryons. The same conclusion can be inferred from the 2D posterior distributions in the $\Omegam$-$S_8$ plane presented in Fig.~\ref{fig:cosmology_hydrosims} for the analyses using the Eagle and the Illustris scenarios as mock data.

\subsection{Information gained with informative prior on baryonic physics}
\label{subsec:informative_baryon_prior}

Next we explore the improvement in the constraints on cosmological parameters when adopting our well-motivated, informative $Q_1$ prior which limits the allowed range of baryonic uncertainties to exclude feedback strength at the Illustris level (\S\ref{subsec:baryon_prior}). Thus, when adopting our informative prior, we do not expect the allowed degrees of freedom to fully mitigate Illustris or other baryonic scenarios with more intense AGN feedback.

As shown in the gray bars of Fig.~\ref{fig:S8_error}, we expect to have about $20\%$ improvement in the marginalized 1D $S_8$ constraint when using the informative $Q_1 \in$ (0,4) prior, compared with adopting Y1 scale cuts (yellow bars). Figure~\ref{fig:cosmology_hydrosims} also provides a visualization of the relative improvements in terms of 2D posterior distributions (gray shaded versus yellow contours).

\subsection{Expected constraints on baryonic parameters}
\label{subsec:baryon_constraint_simulated}

Next we present the expected constraints on the baryon parameters (PC amplitudes) for our baseline pipeline setting and discuss the potential parameter projection effects on their posterior distributions.

\begin{figure*}
\begin{center}
\includegraphics[width=0.96\textwidth]{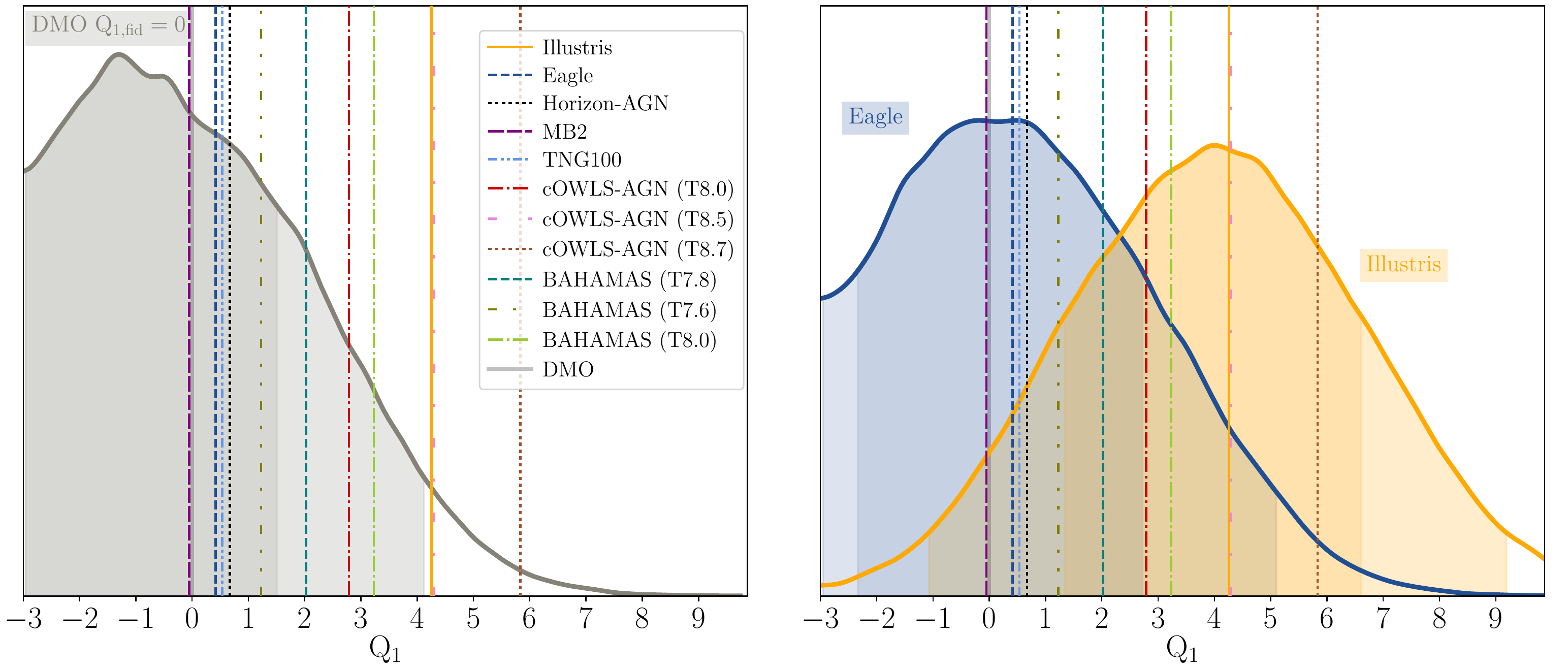}
\caption{Posterior distributions of $Q_1$ under our baseline setting for mock data vectors of DMO (left panel),and Illustris and Eagle scenarios (right panel). We show the case when cosmic shear is extended to 2.5$\arcmin$, and only varying $Q_1$ to account for baryons. 
\textit{Given the constraining power of Y1, when the baryonic feedback of the input mock data is weak (at the level of Eagle), we expect to exclude some of the extreme baryonic scenarios at the $\sim 2\sigma$ confidence intervals.
}} 
\label{fig:Q1dist_sims}
\end{center}
\end{figure*}

Figure~\ref{fig:Q1dist_sims} shows the $Q_1$ posterior distributions for the DMO (left panel) and Eagle and Illustris (right panel) mock data vectors, with cosmic shear data down to 2.5$\arcmin$ and with only $Q_1$ being marginalized over an un-informative prior. The theoretical values of $Q_{1}$ for various baryonic scenarios are computed relying on the relation of Eq.~\eqref{eq:invLB-M}, as detailed in \S\ref{subsec:baryon_prior}. 

The DMO mock data are created with $Q_1=0$, and therefore could be used to estimate the level of projection effects on $Q_1$. We see that the marginalized 1D peak of $Q_1$ has a $\sim 0.5\sigma$ shift from its fiducial value.
This happens due to parameter degeneracies between $Q_1$ and other cosmological and systematics parameters. As discussed in \S\ref{subsec:residual_bias}, we have seen that biases of order $\sim 0.3\sim 0.5\sigma$ are expected (see the yellow square markers in Fig.~\ref{fig:S8_1D}) in the marginalized 1D $S_8$ constraints for the case of DMO. We explore the topic of parameter degeneracies in more detail in Appendix~\ref{sec:parameter_degeneracy}.

Note that the projection effect in $Q_1$ is less apparent for the cases of Eagle ($\sim 0.01\sigma$) and Illustris ($\sim 0.1\sigma$), as shown in the right panel of Fig.~\ref{fig:Q1dist_sims}. This is because the mock data vectors of Eagle and Illustris have extra baryonic features hidden in the higher-order PC modes. This residual ``noise'' of baryonic physics, which can not be accounted for using only $Q_1$, is pushing the $Q_1$ posterior closer to the expected theoretical value.

Regarding the constraining power on baryonic physics shown in Fig.~\ref{fig:Q1dist_sims}, we find that the DES Y1 type constraint can exclude baryonic scenarios that are different from the input fiducial scenario by $\sim 2 \sigma$. For example, when the input fiducial baryonic scenario is weak like Eagle (blue curve in the right panel), the cosmoOWLS-AGN with the minimum heating temperature at $10^{8.7} K$ (the strongest feedback baryonic scenario in the pool) can be excluded at the $2 \sigma$ confidence level, given Y1 statistical power. When the input mock baryonic scenario is Illustris (yellow curve in the right panel), all other baryonic scenarios are covered within the $2\sigma$ posterior region of $Q_1$.

\section{Cosmology Constraints from DES Y1 data}
\label{sec:cosmology_constraint}

This section presents the main cosmology results when applying our pipeline to DES Y1 data. 

\begin{figure*}
  \begin{center}
  \includegraphics[width=0.95\textwidth]{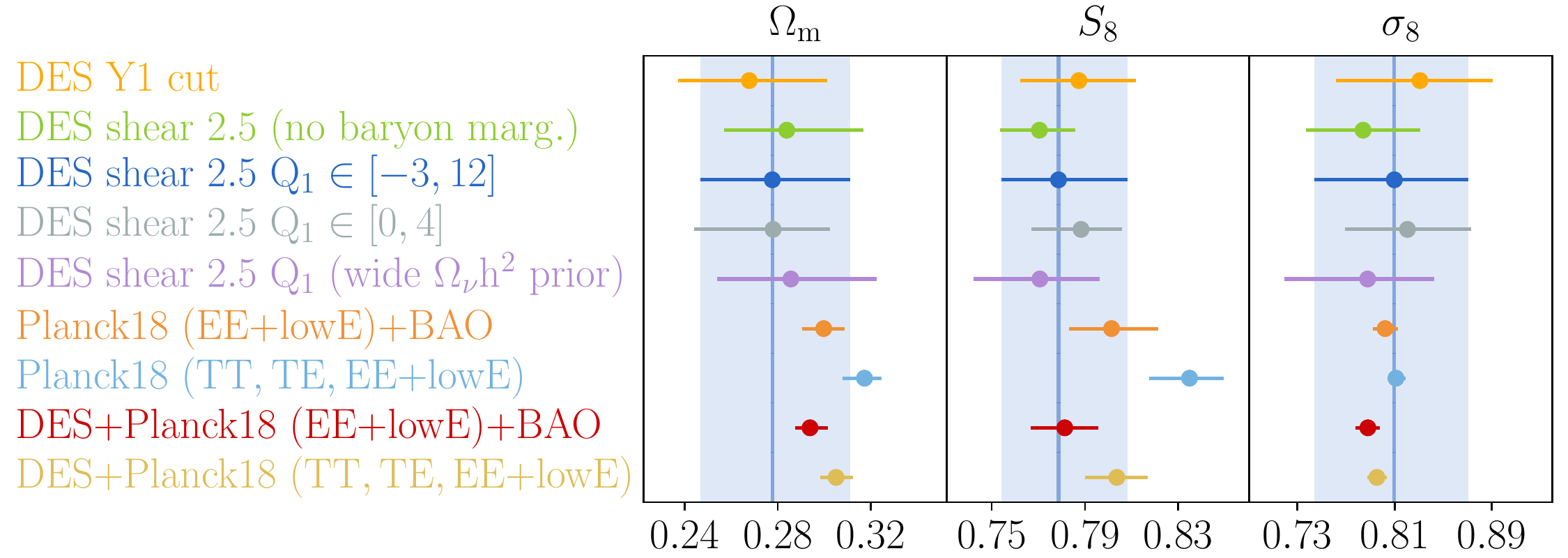}
  \caption{
  Summary of the marginal 1D peak values of cosmological parameters in \LCDM from DES Y1 data. 
  The 68\% confidence levels are shown as horizontal bars.
  The 1st row is the baseline DES 3$\times$2pt data analysis with the adoption of the conservative Y1 scale cuts, and with the informative neutrino mass prior being applied.  
  Rows 2--4 rows present the DES results with cosmic shear extended to 2.5$\arcmin$, but without taking baryonic uncertainty into account (yellow-green), using the $Q_1$ parameter to marginalize over the effect of baryons with an uninformative $Q_1$ prior (blue), and with an informative $Q_1$ prior (gray). The 5th row (purple) shows the result using the same setup as the 2nd row, but adopting wide neutrino prior as in the original DES Y1 analysis.
  Two likelihood chains from the Planck DR18 results \citep{Planck18} are presented for comparison: the CMB polarization auto power spectra combined with BAO (orange), and the joint CMB temperature and polarization auto- and cross-power spectra (light blue). The last two rows are the results of the small-scale extended DES data (as shown in 3rd row) when adopting informative cosmological priors from Planck EE+BAO (red), and from Planck TT,TE,EE (yellow).
 }
  \label{fig:cosmology_parameter_cp}
  \end{center}
\end{figure*}

Figure~\ref{fig:cosmology_parameter_cp} presents a summary of the 68\% confidence intervals on the constraints of $\Omegam$, $S_8$ and $\sigma_8$ for all the analyses we have run. 
As a high-level summary, we start by presenting the DES Y1-only constraints of the baseline setting with Y1 scale cuts applied (top orange), and with the cosmic shear data extended to 2.5\arcmin\ but without performing baryon marginalization (green). With baryonic effects being properly marginalized through an non-informative prior on $Q_1$, we find that almost no information is gained with the inclusion of small-scale cosmic shear data points (dark blue) compared to the case with conservative Y1 scale cuts.
When using our informative prior, we find an improvement in cosmological constraints from the inclusion of small-scale cosmic shear (gray). Finally, we also present a result with cosmic shear being extended down to 2.5\arcmin\ but with an adoption of the non-informative prior on neutrino mass (purple). We then combine the DES Y1 constraints with external data sets of Planck and baryon acoustic oscillation (BAO) constraints (darker orange and light blue), and explore the combined results (red and yellow).
We will discuss these results in detail below.

\subsection{Baseline constraints}

Our baseline setting mostly follows the fiducial DES Y1 3$\times$2pt key paper \citep{DES18}, except for adopting a informative neutrino prior of ${m_{\nu}} < 0.12$ eV/$c^2$ (see \S\ref{subsec:likelihood_analysis}). 

\begin{figure*}
\begin{center}
\includegraphics[width=0.95\textwidth]{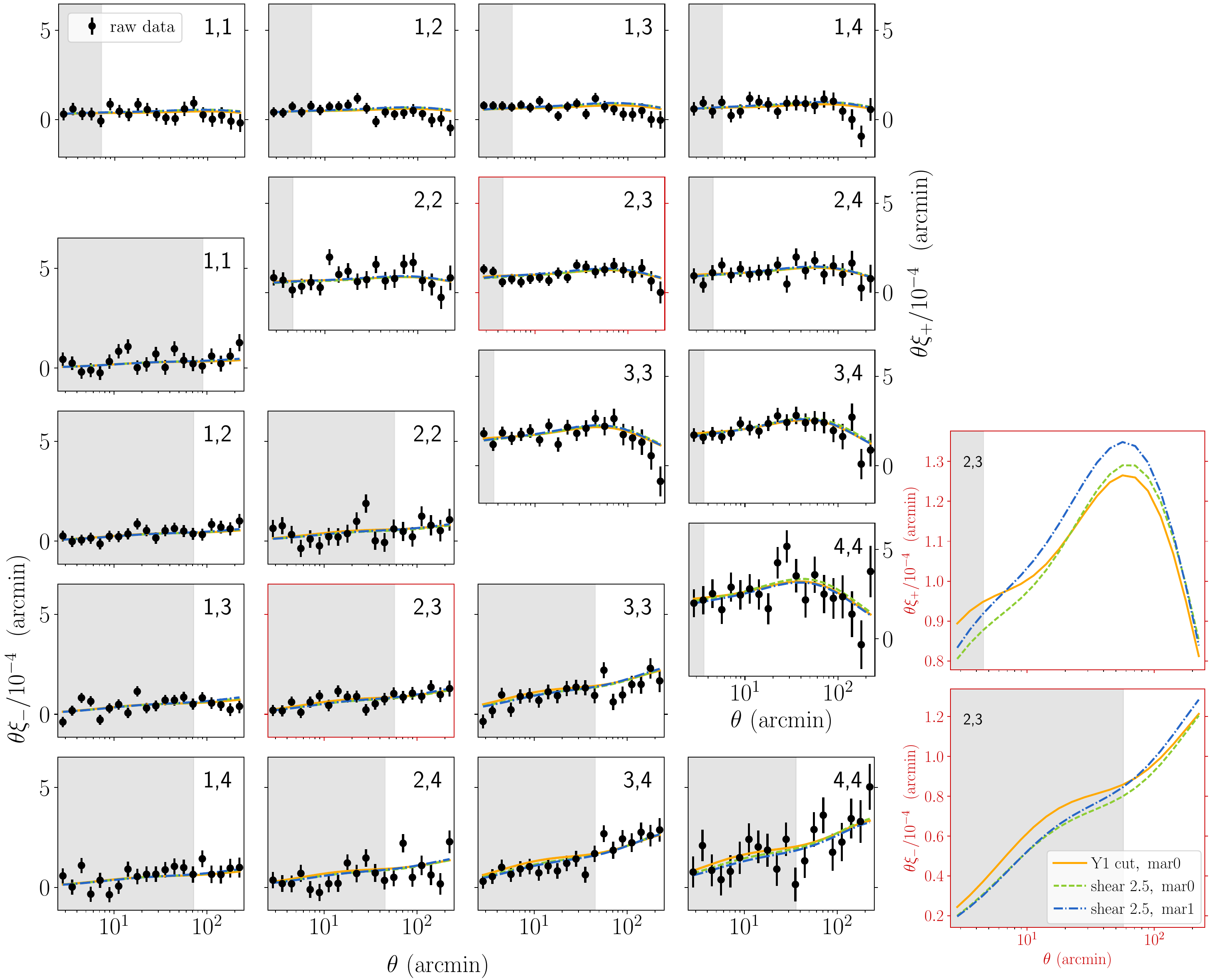}
\caption{DES Y1 cosmic shear data vector (filled black points) and the best-fit theoretical predictions from our baseline analyses. 
The yellow solid lines show the best-fit model when the original Y1 scale cut is applied; the Y1 discarded data points are highlighted in the shaded gray regions. The blue dash-dotted (yellow-green dashed) lines indicate the best-fit model when extending cosmic shear data points to 2.5$\arcmin$, and marginalizing over 1 (0) PC amplitude to account for baryons. For clarity, the two panels on the right show the (2, 3) tomographic bin with the rescaled vertical axis to better illustrate the differences between the models.
The $\chi^2$ information of these best-fit models is summarized in Table~\ref{tb:chi2}.} 
\label{fig:DES_xi_dataV}
\end{center}
\end{figure*}

Figure~\ref{fig:DES_xi_dataV} shows the best-fit theoretical models on top of the observed cosmic shear correlation function. The yellow lines show the fits when the original Y1 scale cut is applied. With the discarded small-scale data points added in the analyses, the yellow-green contours are the result without performing baryon marginalization; the blue lines show the result when the first PC amplitude is used to marginalize over uncertainties in baryonic effects. 
In Table~\ref{tb:chi2}, we provide $\chi^2$ analyses on the derived best-fit models. For the case of Y1 scale cut (first column), the reduced $\chi^2$ derived in this work is consistent with the fiducial DES Y1 key paper (as discussed in the Appendix C of \citealt{DES18}). After including the  extra 175 small-scale data points of cosmic shear, but without introducing any new parameters in the modeling procedure, the reduced $\chi^2$ value remains low (second column of Table~\ref{tb:chi2})\footnote{Note that for the small-scale data analyses, the $\chi^2$ value goes from 674 to 675 when an additional degree of freedom is added to perform baryon marginalization. This could happen due to the stochastic MCMC sampling in high dimensional parameter space, and that the likelihood surface is not a simply smooth function but noisy. So when comparing the rediced $\chi^2$ values, their error bars ($\sqrt{2/d.o.f}$) are important.}. 
This suggests that the baryonic features in the power spectrum on the scales to which these data are sensitive are weak enough that, within the Y1 error, the DMO calibrated theoretical model still provides a valid description of the data. Adding an extra degree of freedom to account for the potential baryonic effect at small scales does not reduce the $\chi^2$ value any further.

\begin{table*}
\caption{Goodness-of-fit for 3$\times$2pt data for the best-fit models (maximum likelihood point sampled in a chain), and the summary of constraints on the 1D peak value of $\Omegam$, $S_8$, and $\sigma_8$. The comparison between the cosmic shear data vectors and the model predictions for these best-fit models are shown in Fig.~\ref{fig:DES_xi_dataV}. The first row lists the $\chi^2$ values. The second and the third rows summarize the effective number of parameters (parameters subjected to wide priors), and the effective degrees of freedom for each of the analyses settings. The fourth row shows the reduced $\chi^2$ values computed, with their errors provided in the fifth row. 
To understand whether a specific model is a good description of the data, in the sixth row, we derive the $p$-values based on the $\chi^2$ distribution. A $p$-value > 0.05 indicates that the data are compatible with the model prediction within the error.}
\begin{center}
\begin{tabular}{lcccc}
\hline
		  	&     Y1 cut    	& 	\makecell{ shear to 2.5\arcmin \\ (no baryon marginalization) }	&	\makecell{ shear to 2.5\arcmin \\ (mar. $Q_1$) }	& \makecell{ DES (shear 2.5\arcmin, mar. $Q_1$) \\ $+$ \\ Planck EE+BAO } \\  \hline
best-fit $\chi^2$			&	502		&	674				&	675			& 678	\\
effective $N_{\rm par}$	& 	12		& 	12				&	13	& 8			\\
effective d.o.f			& 	443		&	618				& 	617			& 622	\\
reduced $\chi^2$		& 	1.128	&	1.091			&	1.094			& 1.090 \\
$\sqrt{2 / {\rm d.o.f}}$	& 	0.067	& 	0.057			& 	0.057		&	0.057 \\
$p$-value				& 	0.027	& 	0.059			&	0.053		& 0.059	\\
\hline \hline
$\Omegam$  &   $0.268^{+0.034}_{-0.031}$ & $0.284^{+0.033}_{-0.027}$ & $0.278^{+0.034}_{-0.031}$ & $0.294^{+0.008}_{-0.006}$ \\
$S_8$      &   $0.787^{+0.024}_{-0.025}$ & $0.770^{+0.015}_{-0.017}$ & $0.779^{+0.030}_{-0.025}$ & $0.781^{+0.014}_{-0.015}$ \\
$\sigma_8$ &   $0.831^{+0.060}_{-0.069}$ & $0.784^{+0.047}_{-0.047}$ & $0.810^{+0.061}_{-0.066}$ & $0.788^{+0.010}_{-0.010}$\\
\hline

\label{tb:chi2}
\end{tabular}
\end{center}
\end{table*}

\begin{figure*}
\begin{center}
\includegraphics[width=0.98\textwidth]{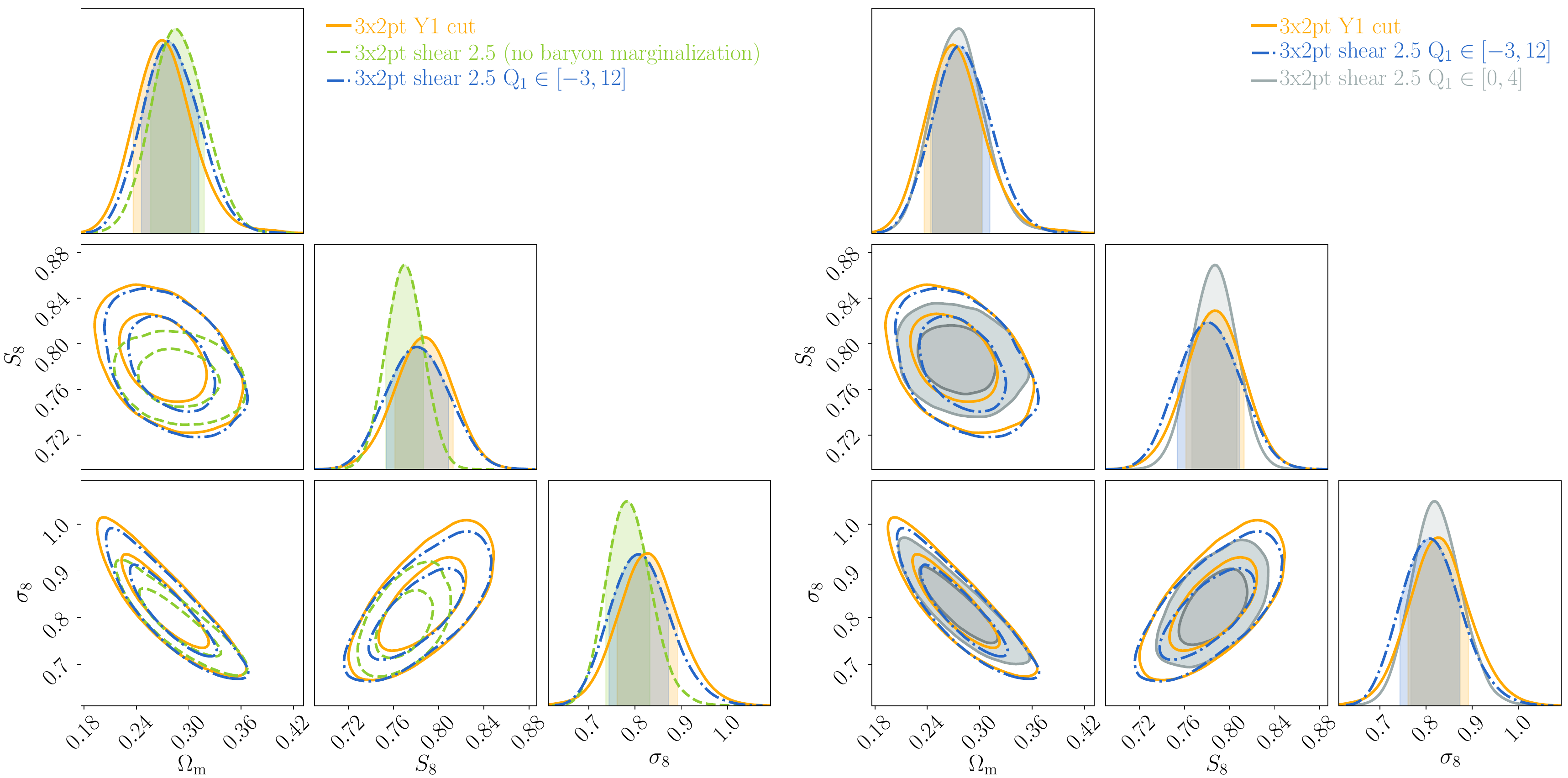}
\caption{Cosmological parameter constraints on DES Y1 3$\times$2pt data in \LCDM. The yellow solid contours indicate the constraint when the original Y1 scale cut is applied. The blue dash-dotted contours show the result when adding extra small-scale cosmic shear data, and with uncertainties of baryon being marginalized with non-informative $Q_1$ prior.
[Left panel] The yellow-green dashed contours indicate the result with small-scale cosmic shear data added but without perfroming baryon marginalization. [Right panel] The shaded gray contours indicate the case when extending cosmic shear to 2.5\arcmin\ while adopting informative prior on the first PC amplitude to perform baryon marginalization. 
} 
\label{fig:cosmology_DESY1}
\end{center}
\end{figure*}

The posterior cosmological parameter distributions from our baseline analyses are presented in Fig.~\ref{fig:cosmology_DESY1}. The yellow contours show the constraint when the original Y1 scale cuts are applied.
The derived marginalized 1D peak constraints are
\begin{equation}
\label{eq:cosmology_Y1cut}
\Omegam = 0.268^{+0.034}_{-0.031} \quad S_8 = 0.787^{+0.024}_{-0.025} \quad \sigma_8 = 0.831^{+0.060}_{-0.069}\ ,
\end{equation}
which is consistent with the Y1 key paper result. The minor difference in the constraints is caused by the following three factors: neutrino prior difference (which would induce a factor of $\sim 0.5\sigma$ shift as shown in Fig.~\ref{fig:neutrino_prior}), sampler difference (\texttt{emcee} v.s. \texttt{MultiNest} -- \citealt{Feroz09}) and theory code uncertainty (\textsc{CosmoLike} v.s. \textsc{CosmoSIS} -- \citealt{Zuntz15}). As discussed in Fig.~17 of \citealt{DES18}, the latter two differences together contribute at about $0.2\sigma$ level. Given the Y1 constraining power, these uncertainties would not change the conclusion of this paper, but further investigation and management of their error budgets will become a necessity for Stage IV cosmological analyses.

The blue contours in Fig.~\ref{fig:cosmology_DESY1} show the cosmological constraints when introducing additional information from small-scale cosmic shear in the original Y1 3$\times$2pt analysis while properly marginalizing over the uncertainties of baryons at these scales. The derived marginalized 1D posterior peaks of the main cosmological parameters are
\begin{equation}
  \label{eq:cosmology_smallQ1}
  \begin{split}
  \Omegam & = 0.278^{+0.034}_{-0.031} \\ 
  S_8 & = 0.779^{+0.030}_{-0.025} \\ 
  \sigma_8 & = 0.810^{+0.061}_{-0.066}\ .
  \end{split}
\end{equation}

Without performing baryon marginalization (yellow-green contours), the resulting marginal 2D posterior is still overlapping with the $1\sigma$ region of the case when introducing one parameter to account for baryonic uncertainty (blue), and with the result when the conservative scale cuts is adopted (yellow). By comparing the $\Omegam$-$S_8$ posterior distribution in the left panel of Fig.~\ref{fig:cosmology_DESY1} with the left panel of Fig.~\ref{fig:cosmology_hydrosims}, we find that the baryonic scenario as measured from DES is comparable to that of the Eagle mock-data simulation, for which we have learned that even without performing baryon marginalization when extending cosmic shear to 2.5$\arcmin$, the $S_8$ bias can still be within 0.5$\sigma$ (see \S\ref{subsec:residual_bias}).

\subsection{Informative prior on baryonic physics}

The parameter constraints obtained when adopting our informative $Q_1$ prior is shown in the right panel of Fig.~\ref{fig:cosmology_DESY1} in the shaded gray contours in comparison with the baseline (blue contours) and the Y1 scale cut (yellow contours) results. The derived marginal 1D peak cosmological parameters are
\begin{equation}
  \label{eq:cosmology_smallQ1_info3}
  \begin{split}
  \Omegam & = 0.278^{+0.024}_{-0.034} \\ 
  S_8 & = 0.788^{+0.018}_{-0.021} \\ 
  \sigma_8 & = 0.821^{+0.052}_{-0.052}\ .
  \end{split}
\end{equation} Compared with the (averaged) error bars resulting from setting Y1 scale cuts, as shown in Eq.~\eqref{eq:cosmology_Y1cut}, we have around 11\%, 20\%, 19\% improvements on the 1D marginalized $1\sigma$ error bars of $\Omegam$, $S_8$, $\sigma_8$, respectively. The derived improvements are consistent with what we have learned from simulated likelihood analyses (\S\ref{subsec:informative_baryon_prior}) before unblinding.

\subsection{Non-informative neutrino prior}

\begin{figure}
  \begin{center}
  \includegraphics[width=0.47\textwidth]{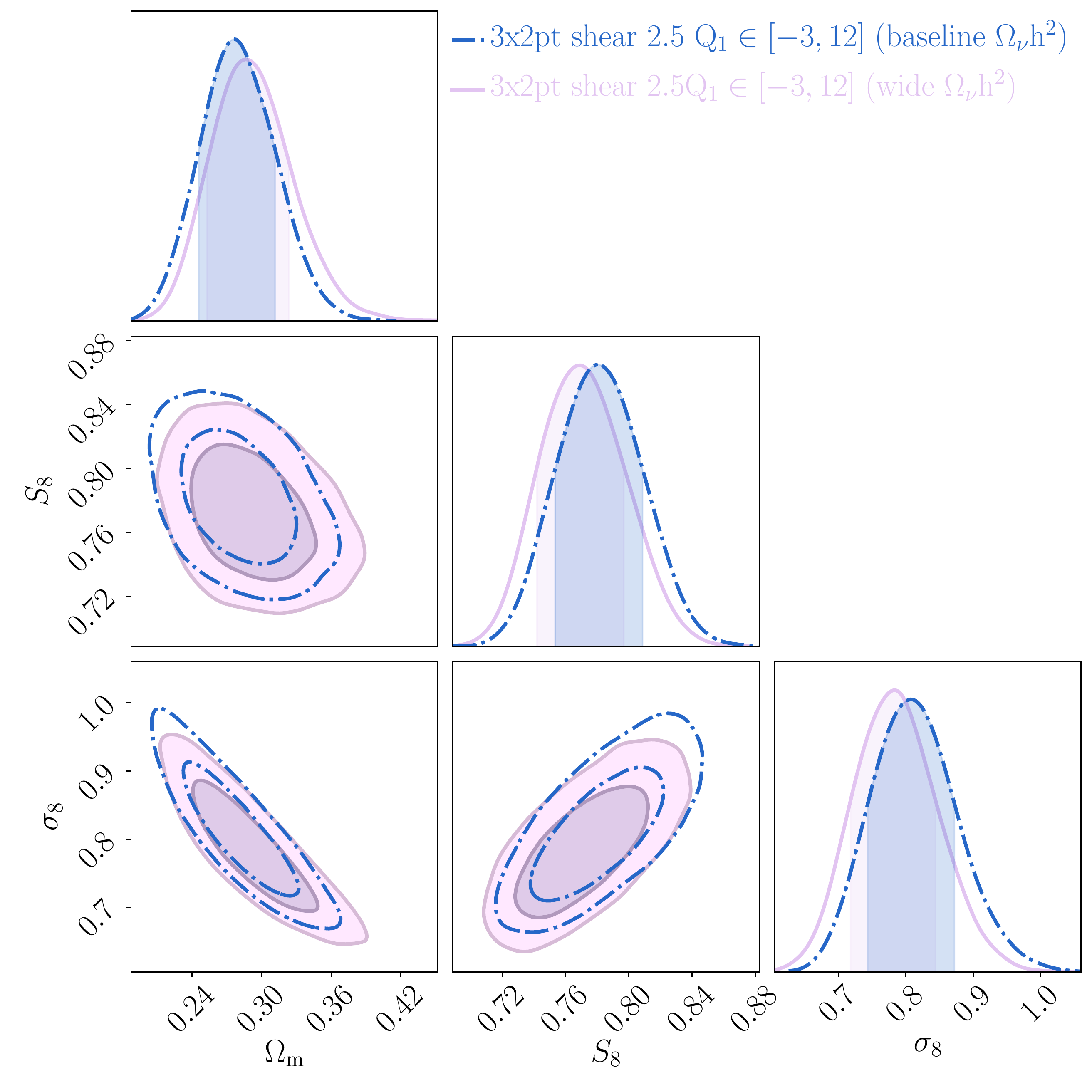}
  \caption{ The effect of neutrino priors on the cosmological parameter constraints with DES Y1 3$\times$2pt data in \LCDM.
  The blue contours show our baseline result as in Fig.~\ref{fig:cosmology_DESY1}, for which an informative neutrino prior is applied ($\Omega_{\nu}h^2 \in$ Flat($5\times10^{-4}, 1.3\times10^{-3}$)). The shaded purple contours indicate the result when adopting a non-informative prior on neutrinos as the original Y1 analysis ($\Omega_{\nu}h^2 \in$ Flat($5\times10^{-4}, 0.1$)).
  }
  \label{fig:cosmology_DESY1_neutrino}
  \end{center}
\end{figure}

Due to concerns about parameter projection effects as discussed in \S\ref{subsec:neutrino_prior}, we adopt an informative prior on the sum of the neutrino mass parameter based on the external information from \citet{Planck18} and BAO measurements \citep{Beutler11, Ross15, Alam17}.
Here we explore the cosmology results for different choices of neutrino priors between our baseline ($\Omega_{\nu}h^2 \in$ Flat($5\times10^{-4}, 1.3\times10^{-3}$)) and the non-informative case as adopted in the original Y1 analysis ($\Omega_{\nu}h^2 \in$ Flat($5\times10^{-4}, 0.1$)). 

Figure~\ref{fig:cosmology_DESY1_neutrino} shows that the adoption of non-informative neutrino priors (purple contours) results in a slight shift ($\sim0.3\sigma$ in $S_8$) in the posterior distribution compared to the case of an informative neutrino prior, which matches our previous observation using simulated likelihood analyses (see \S\ref{subsec:neutrino_prior}) before unblinding. 
The best-fitting cosmological parameters when including cosmic shear small-scale information and when marginalzing over uncertainties in baryonic physics with a non-informative prior on the $Q_1$ parameter are as follows:
\begin{equation}
  \label{eq:cosmology_smallQ1_info}
  \begin{split}
  \Omegam & = 0.286^{+0.037}_{-0.032} \\ 
  S_8 & = 0.771^{+0.026}_{-0.028} \\ 
  \sigma_8 & = 0.788^{+0.055}_{-0.069}\ .
  \end{split}
\end{equation}

\subsection{Constraints with external data}
\label{subsec:cosmlogy_external}

We compare our baseline DES measurements to external data from Planck \citep{Planck18} and BAO measurements \citep{Beutler11, Ross15, Alam17}. 
The main motivation is to use external information to tighten constraints on cosmology, and increase our constraining power on baryonic physics (see \S\ref{sec:baryon_constraint}).

\begin{figure*}
  \begin{center}
  \includegraphics[width=0.98\textwidth]{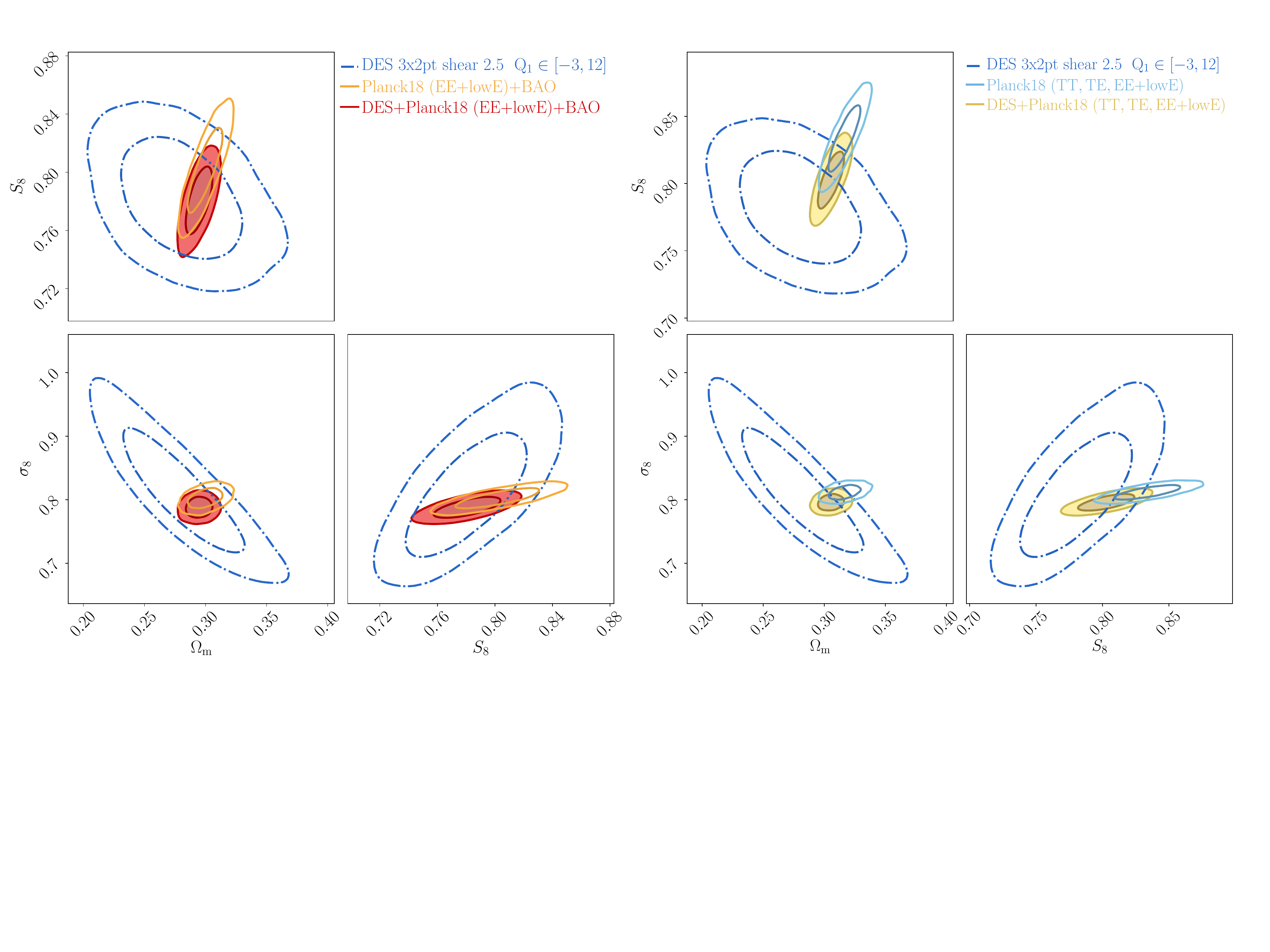}
  \caption{DES+Planck \LCDM cosmology constraints. 
  The blue contours show our baseline analysis on DES 3$\times$2pt data with small-scale cosmic shear data included in the analysis, and marginalizing over the $Q_1$ parameter to account for baryon uncertainty. 
  The orange contours in the left panel show the Planck EE+lowE+BAO constraints, and the light blue contours in the right panel display the Planck TT, TE, EE+lowE results from \citet{Planck18}. Within $2\sigma$, the Planck contours are in agreement with our baseline DES result. The shaded contours present the joint constraints from our baseline DES analyses with the information from the Planck results.}
  \label{fig:cosmology_DESxPlanck}
  \end{center}
\end{figure*}

We have considered two likelihood chains from the baseline DR18 Planck analyses\footnote{\href{https://wiki.cosmos.esa.int/planck-legacy-archive/index.php/Cosmological_Parameters}
{2018 Planck Cosmological parameters and MCMC chains}}: the CMB polarization auto power spectra combined with BAO (referred to as EE+lowE+BAO), and the joint CMB temperature and polarization auto- and cross-power spectra (referred to as TT, TE, EE+lowE).

Our primary choice is the Planck EE+BAO likelihood, motivated by its high level of statistical consistency with DES Y1, as shown in the left panel of Fig.~\ref{fig:cosmology_DESxPlanck}. 
We compute 5-dimensional parameter covariance in $\Omegam$, $A_s$, $n_{\rm s}$, $\Omegab$, and $h$ from the Planck EE+BAO posterior distribution, and then rerun the DES Y1 data by adopting informative 5-dimensional Gaussian priors on these cosmological parameters. We have confirmed that the posterior of the Planck chains can be well-described with a multidimensional Gaussian fit out to the $4\sigma$ level.
The shaded red contours in Fig.~\ref{fig:cosmology_DESxPlanck} present the combined result. The $\chi^2$ analysis on the sampled maximum likelihood model indicates that our model prediction is consistent with the data (see Table~\ref{tb:chi2}).

The 1D marginal constraints are
\begin{equation}
  \label{eq:cosmology_smallQ1_info2}
  \begin{split}
  \Omegam & = 0.294^{+0.008}_{-0.006} \\ 
  S_8 & = 0.781^{+0.014}_{-0.015} \\ 
  \sigma_8 & = 0.788^{+0.010}_{-0.010}\ .
  \end{split}
\end{equation} There is $\sim  50\%$ improvement in the $S_8$ constraint after adopting informative cosmological priors.

We also compare our DES Y1 analysis with the Planck TT, TT, TE constraint (light blue contours of Fig.~\ref{fig:cosmology_DESxPlanck}). With the addition of Planck CMB temperature information, the CMB constraints reveal hints of tension with several ongoing weak lensing experiments \citep{Hildebrandt17, Hikage19, DES18}, where the weak lensing results show lower values in the $S_8$ constraints. As shown in the right panel of Fig.~\ref{fig:cosmology_DESxPlanck}, although the two datasets are largely in agreement to within the $95\%$ confidence level in their 2D posterior constraints, the marginal 1D $S_8$ constraints differ by more than $1\sigma$ (see summary in Fig.~\ref{fig:cosmology_parameter_cp}).

\section{Baryon Constraints from DES Y1 data}
\label{sec:baryon_constraint}

In this section we present the constraints on baryonic physics in terms of the first PC amplitude $Q_1$, which captures the most dominant features of baryonic effects on the clustering of the matter distribution. We first discuss the baryonic physics constraints from DES alone, and then increase the constraining power by combining DES with external data from Planck and BAO measurements.

\begin{figure}
  \begin{center}
  \includegraphics[width=0.48\textwidth]{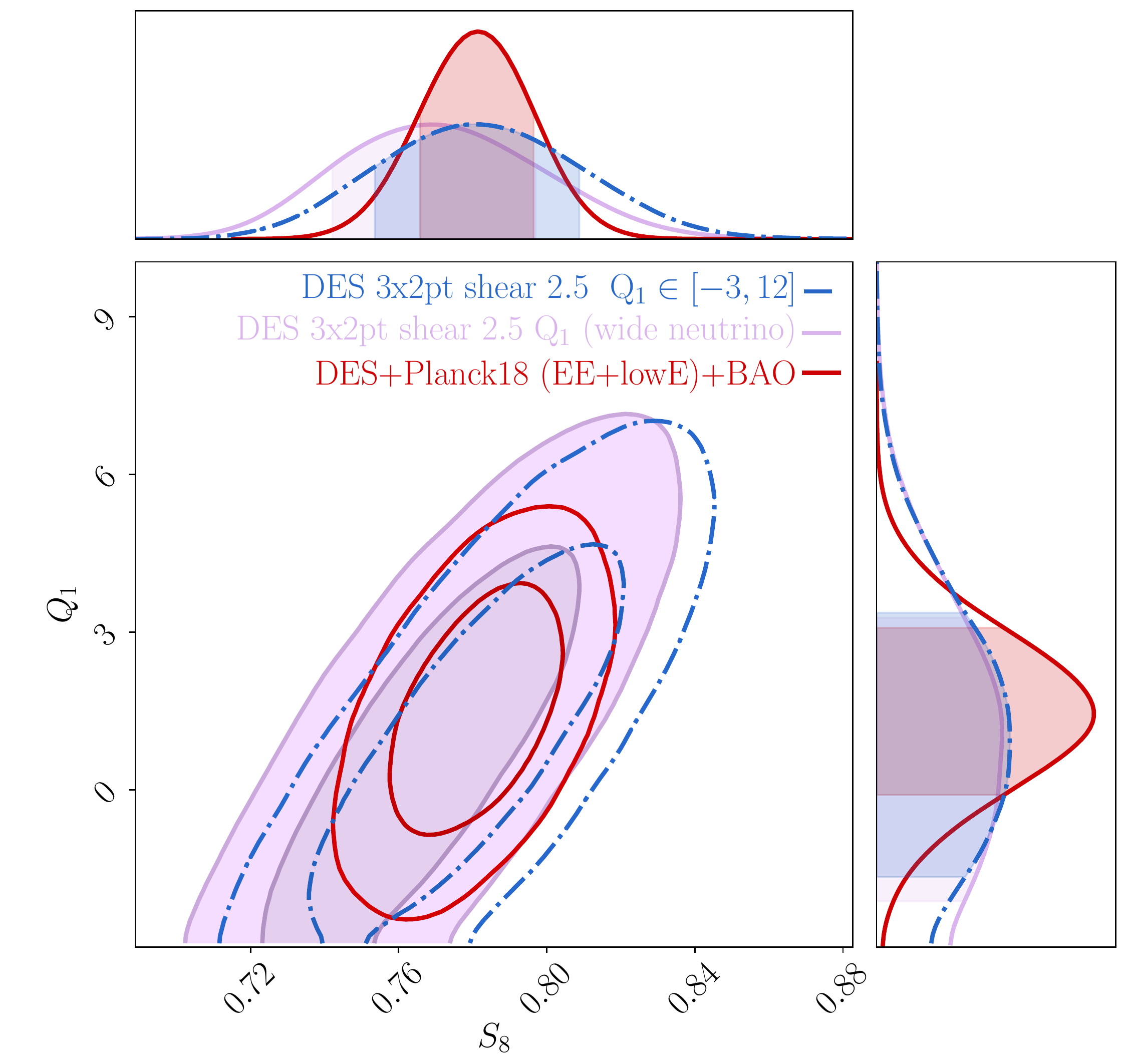}
  \caption{Joint constraints on cosmological parameters and baryonic physics. The blue contours show the $Q_1$--$S_8$ constraints from the baseline analyses of DES 3$\times$2pt data with cosmic shear measured down to 2.5\arcmin\ and with the baseline priors detailed in Table~\ref{tb:params}. 
  The analysis setting for the purple contour is the same as the blue contour, except for the adoption of the non-informative neutrino prior as used in the original Y1 analysis. 
  The red contours present the results when adopting informative cosmology priors from the external information of Planck 2018 EE+BAO.}
  \label{fig:DES_Q1_S8}
  \end{center}
\end{figure}

\begin{figure*}
  \begin{center}
  \includegraphics[width=0.65\textwidth]{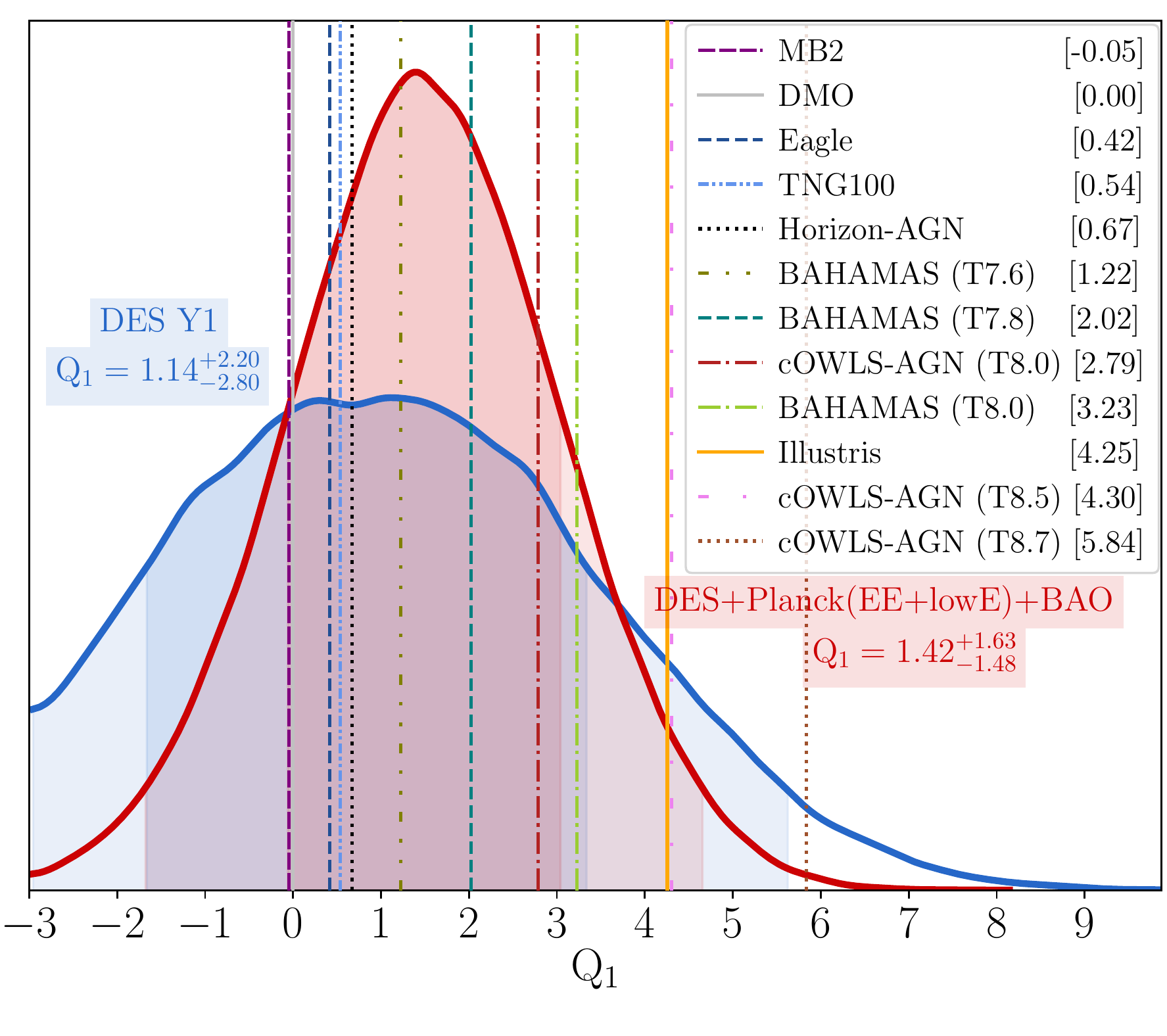}
  \caption{ Constraints on the baryonic parameter $Q_1$ from the DES Y1 information only (blue) and with the adoption of the Planck EE+BAO cosmological prior (red). 
  The shaded bands highlight the 68\% and the 95\% confidence intervals of the constraints. The representative $Q_1$ values from various baryonic scenarios are shown in the legend and are over-plotted as vertical lines.
\textit{The most extreme baryonic scenario, the cosmoOWLS-AGN with minimum AGN heating temperature of $10^{8.7} K$, is excluded at above $2\sigma$ level by the real data.}
  }
  \label{fig:DES_Q1_2sigma}
  \end{center}
  \end{figure*}

\subsection{DES only information}

The blue curves shown in Fig.~\ref{fig:DES_Q1_S8} indicate the joint constraints of $Q_1$ and $S_8$ from the DES 3$\times$2pt data with cosmic shear extended down to 2.5$\arcmin$, and with parameter priors listed in Table~\ref{tb:params}. As discussed in Appendix~\ref{sec:parameter_degeneracy}, there is a significant positive correlation between $Q_1$ and $S_8$. 

The marginal 1D $Q_1$ posterior for our baseline result is presented in the blue curve of Fig.~\ref{fig:DES_Q1_2sigma}.
We find that $Q_1$ is constrained to be in the range: 
\begin{equation}
  \label{eq:Q1_DESY1}
  \begin{split}
   -1.66 < Q_1 < 3.34 & \mbox{\ \ \ (68\%,\ \ DES Y1)} \\ 
   -2.96 < Q_1 < 5.63 & \mbox{\ \ \ (95\%,\ \ DES Y1)} \ .
  \end{split}
\end{equation} 
We can rule out the cosmo-OWLS scenario with AGN minimum heating temperature setting at $10^{8.7}$ K at $\sim 2.1\sigma$ with DES alone.
This conclusion still holds (and the $Q_1$ posterior remains similar) for the analysis when adopting the original Y1 wide neutrino prior, as shown in the purple contours of Fig.~\ref{fig:DES_Q1_S8}.

As concluded in \S\ref{subsec:baryon_constraint_simulated} from the results of the simulated likelihood analyses, we expect a $\lesssim 0.5\sigma$ shift in the peak of the marginal 1D $Q_1$ posterior distribution due to the parameter projection effects driven by the degeneracies between parameters.

In the next section, we explore the constraints on baryonic physics when including cosmological information from external datasets. 

\subsection{Adopting cosmological parameter priors from external datasets}

As discussed in \S\ref{subsec:cosmlogy_external}, we adopt the Planck EE+BAO likelihood as the primary source of prior information on cosmological parameters, due to its consistency with DES Y1 data.

With the inclusion of the Planck EE+BAO information, the constraints on cosmology improve by $\sim 38\%$ in the $68\%$ confidence interval of $Q_1$ (see the red curve in Fig.~\ref{fig:DES_Q1_2sigma}), which is further quantified as:
\begin{equation}
  \label{eq:Q1_DESxPlanckEEBAO}
  \begin{split}
   -0.06 < Q_1 < 3.04 & \mbox{\ \ \ (68\%, \ \ DES+Planck\ EE+BAO)} \\ 
   -1.68 < Q_1 < 4.66 & \mbox{\ \ \ (95\%, \ \ DES+Planck\ EE+BAO)} \ .
  \end{split}
\end{equation} 
With the tighter constraining power, the cosmo-OWLS $10^{8.7}$ K scenario is disfavored by $\sim 2.8\sigma$.

\begin{figure*}
  \begin{center}
  \includegraphics[width=0.99\textwidth]{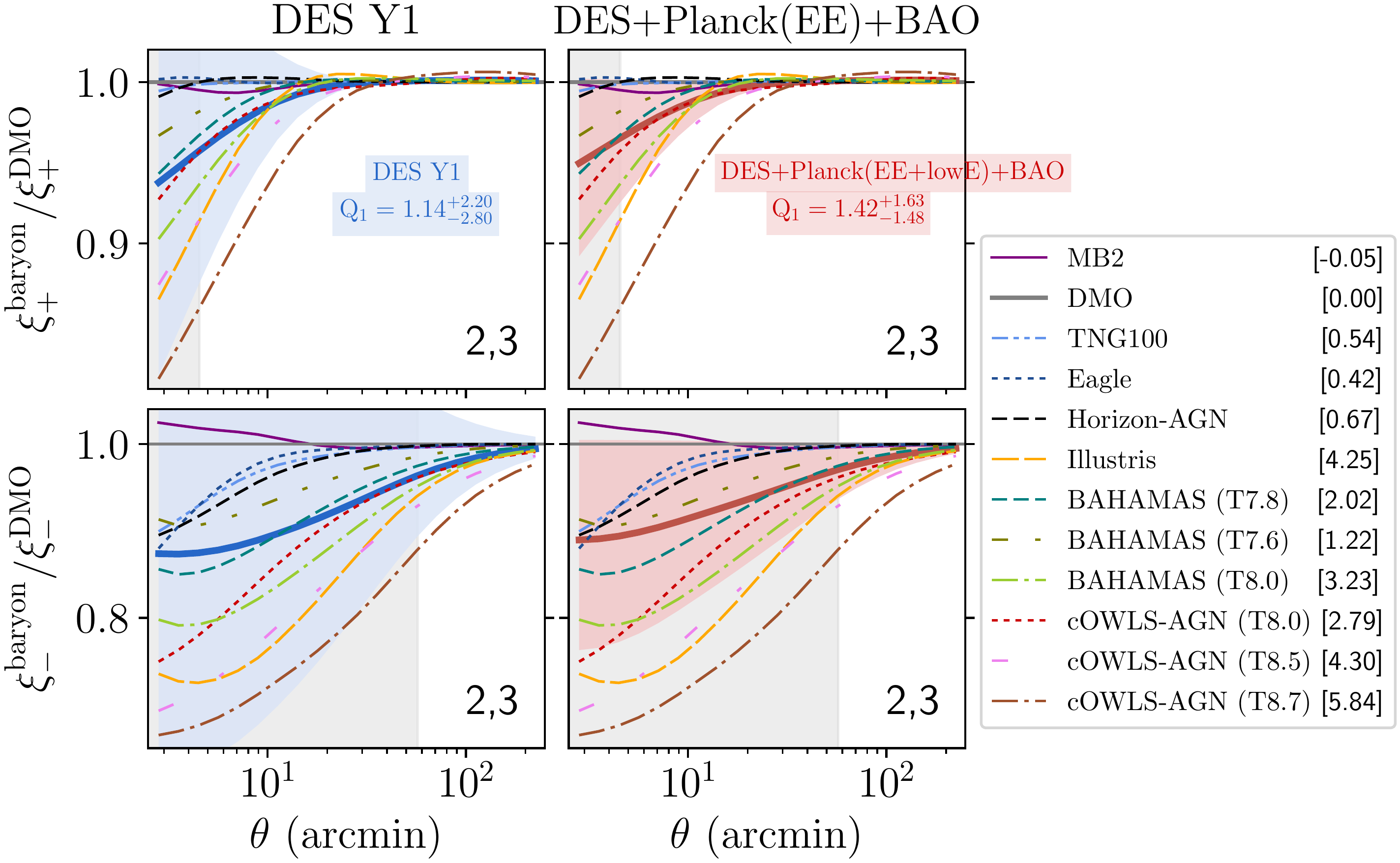}
  \caption{Quantifying the strength of baryonic feedback using the $Q_1$ parameter. Here we show the ratio of cosmic shear correlation functions for the 11 baryonic scenarios with respect to the theoretical (DMO) predictions (thinner curves), using the pair of tomographic bins (2,3) as an demonstration. The gray bands highlight the angular scales that were excluded in the fiducial DES Y1 analysis, but are now included in this work. 
  The thick curves depict the baryonic features using the best-fit $Q_1$ constraints from our baseline DES Y1 analysis (left panel, blue) and the result with the adoption of cosmological priors from Planck 2018 EE+BAO information (right panel, red). The shaded bands highlight the 1$\sigma$ region of our data constraints. The numbers in the colored-shaded legend are the best-fit $Q_1$ values from data constraints, and the representative $Q_1$ values from various baryonic scenarios are also provided in the right-hand-side legend.
  }
  \label{fig:ratio_DESdatav}
  \end{center}
\end{figure*}

We note that when claiming the cosmo-OWLS ($10^{8.7}$ K) scenario is ruled out at $>2\sigma$ significance with DES Y1 data, this does not mean that we are ruling out setting the subgrid physical parameter of $\Delta T_{\rm heat}$ at $10^{8.7}$ K when running the hydrodynamical simulations at $>2\sigma$. Rather, we are simply ruling out the level of baryonic suppression in the cosmic shear signal exhibited by that simulation,  as shown in Fig.~\ref{fig:ratio_DESdatav} based on the empirical PCA framework. As discussed in \S\ref{subsec:mock_data}, for baryon models with realistic physical parameterizations, such as the baryon correction model of \citet{Schneider20} or subgrid physical parameters in hydrodynamical simulations, the baryonic physics parameters controlling halo density profiles couple to the cosmic baryon fraction $f_{\rm b} = \Omegab/\Omegam$. As revealed in the trend of power spectrum ratios in Fig.~6 of \citet{vanDaalen20}, at fixed $\Delta T_{\rm heat}$ value in hydro-simulations, the baryonic  suppression features also vary slightly at different cosmologies with different $f_{\rm b}$. For a universe with lower $f_{\rm b}$, higher AGN heating temperature is needed to generate the same amount of suppression feature compared with a universe with higher $f_{\rm b}$.

Finally, we link back to the physical effect of $Q_1$, which is best demonstrated by looking at the suppression of the amplitude of cosmic shear correlation functions (Fig.~\ref{fig:ratio_DESdatav}, also see Fig.~\ref{fig:Mpar_1sigma_sym}). 
In Fig.~\ref{fig:ratio_DESdatav} we convert the $1\sigma$ constraints on $Q_1$ to cosmic shear model vectors via Eq.~\eqref{eq:Mbary}, and present the ratio of the baryonic physics-included model with respect to the DMO-based theoretical model. We use the pair of tomographic bins (2, 3) to demonstrate the effects of baryons. The thick lines indicate the results when setting the $Q_1$ value at the 1D marginal peak the posteriors, as indicated in the text in the right panel of Fig.~\ref{fig:DES_Q1_S8}. The other baryonic scenarios are also overplotted in thinner curves for comparison. 
The figure shows the effect of the baryonic effects on the shear-shear observables, and can be compared to Fig~\ref{fig:Pk_Ratio} where we show the effects on the matter power spectrum. The shear-shear correlation function measured on a range of scales and tomographic bins can constrain both the spatial and temporal evolution of the baryonic effects.

\section{Discussion and Summary}
\label{sec:summary}

Small-scale information in galaxy imaging surveys has substantial statistical power to improve cosmological constraints. But conventional cosmological analyses discard this information to avoid biased inference of cosmology due to an insufficient theoretical description of the sources of astrophysical and observational systematic uncertainty on these scales.

  The effects of baryonic physics constitute the dominant source of uncertainty at small scales in the matter power spectrum \citep{vanDaalen11, vanDaalen20}.  
  A variety of modeling and mitigation strategies have been proposed in the literature to account for the complicated mechanisms involved, such as baryonic feedback and cooling processes (see \citealt{Chisari19} for a review of existing baryon mitigation methods).

  To enable robust inference of cosmological parameters, it is desirable to find a minimal parameterization that accurately captures the effects of baryonic physics on the observables and to have stringent priors on these parameters. Principal component analysis (PCA) of the cosmological observables, derived from a set of hydrodynamical simulations that span the range of allowed baryonic scenarios, is one of the most promising avenues to obtain such a minimal parameterization \citep{Eifler15, Kitching16, Mohammed18, Huang19}.

  In this paper we employ the hydrodynamical simulation-based PCA method to parameterize baryonic effects in the Dark Energy Survey Year 1 data. We find that one principal component is sufficient to capture the range of baryonic physics at the level of DES Y1 statistical constraining power. We include the amplitude of this PC, $Q_1$, as an additional parameter in our likelihood analysis. 
  The magnitude of $Q_1$ reflects the strength of baryonic feedback, with larger $Q_1$ values corresponding to a stronger suppression of small-scale cosmic shear correlation functions (see Fig.~\ref{fig:Mpar_1sigma_sym}).

Previous DES multi-probe analyses \citep[e.g.,][]{DES18, DES5x2} impose stringent scale cuts to ensure that the analysis is unaffected by baryonic physics; this had the largest effect on the range of scales used by cosmic shear. The inclusion of baryonic effects in our theoretical model for the observables allows us to relax these scale cuts and to include scales as small as 2.5\arcmin\ from cosmic shear in the analysis. We otherwise follow the DES Year 1 systematics modeling and mitigation strategy, except for adding an informative neutrino mass prior based on findings by the Planck satellite mission. The reduced range in varying neutrino mass avoids parameter volume effects in the DES analysis (see \S\ref{subsec:neutrino_prior}).

Our joint analysis of baryonic physics and cosmology (in combination with the other DES systematics parameters) yields $S_8=0.779^{+0.030}_{-0.025}$ if we allow for self-calibration of $Q_1$. When we restrict the range of $Q_1$ such that AGN feedback stronger than the level of Illustris \citep{Vogelsberger14,Genel14} is excluded, we get $S_8=0.788^{+0.018}_{-0.021}$ (see the right panel of Fig.~\ref{fig:cosmology_DESY1}).

We proceed to combine DES Y1 with data from the latest Planck mission analysis \citep{Planck18}. However, we exclude the Planck temperature power spectrum information due to an abundance of caution as to whether these data sets might be in tension \citep{Adhikari19, Park19, Garcia-Quintero19}. 
We instead use the Planck EE+lowE+BAO chain as described in \citet{Planck18}, which also includes BAO measurements from the BOSS DR12, 6DF and MGS survey. Our joint DES Y1+Planck+BAO analysis yields $S_8=0.781^{+0.014}_{-0.015}$ (see Fig.~\ref{fig:cosmology_DESxPlanck}).

We emphasize that the main goal of this paper is not to find the tightest possible constraints on cosmological parameters, but rather unbiased constraints on baryonic physics with cosmological parameters being allowed to vary. We find the baryon parameter $Q_1=1.14^{+2.20}_{-2.80}$ for DES Y1 only and $Q_1=1.42^{+1.63}_{-1.48}$ for DES+Planck EE+BAO (see Fig.~\ref{fig:DES_Q1_2sigma}), which allows us to exclude one of the most extreme AGN feedback hydrodynamical scenario, cosmo-OWLS AGN (T=$10^{8.7}$K), at $\sim2.8\sigma$.

Among the 11 hydrodynamical scenarios in our pool, the default BAHAMAS simulation (minimum AGN heating temperature at T=$10^{7.8}$K) is perhaps the best-calibrated baryon scenario. Not only is it tuned to reproduce the galaxy stellar mass function, but it also has adjusted feedback parameters so that the halo hot gas mass fractions match those from the observations \citep{McCarthy17}.
The $1\sigma$ region of our $Q_1$ posterior constraint likewise includes the default BAHAMAS scenario.

Constraining the strength of baryon feedback is also important to understand whether it can serve as a possible explanation for the ``lensing-is-low'' effect, i.e., the fact that the observed galaxy-galaxy lensing signal is low by $\sim$20-40\% compared to predictions from N-body+HOD mocks, at fixed clustering signal \citep{Leauthaud17}. 
As discussed in \citet{Lange19}, the IllustrisTNG scenario can account for $\sim$10\% of the suppression signal and the stronger feedback scenario of Illustris can reach to $\sim$15\% (c.f. Fig.~\ref{fig:DES_Q1_2sigma} for our constraints on these scenarios). While our constraints currently lack the constraining power to make definite statements on ruling out baryonic effects as a potential explanation for the lensing-is-low signal, we expect that future analyses, e.g., using DES Y3 data, will be very interesting in that regard.

Generally speaking, our resulting $Q_1$ posterior distribution indicates a preference for moderate to weak baryonic feedback, which is consistent with previous cosmic shear constraints of \citet{Joudaki17} for an analysis on CFHTLenS data, and with \citet{MacCrann17} for the DES SV data \citep{DES16SVshear} using \texttt{HMcode} \citep{Mead15}. 
Recent constraints from the cosmic shear KiDS-VIKING 450${^2}$ degree field \citep{Hildebrandt20} analyzed by \citet{Yoon20} also derive a weak signal of baryonic feedback based on the baryonic model of \texttt{HMcode}. They found that the baryon suppression signal is consistent with the DMO scenario within 1.2$\sigma$ significance with the KiDS data alone, and is at 2.2$\sigma$ level deviation from DMO under the assumption of the WMAP9 cosmology.
On the contrary, based on the analysis of the DLS (Deep Lens Survey, \citealt{Jee16}) Fourier space galaxy-mass and galaxy-galaxy power spectra, \citet{Yoon19} report a preference for strong baryonic feedback that is more extreme than that predicted by OWLS-AGN\footnote{The OWLS-AGN scenario is equivalent to the cosmoOWLS AGN scenario with T=$10^{8.0}$K as indicated in the red line of Fig.~\ref{fig:DES_Q1_2sigma}}. 
The difference in the resulting baryon constraints could be the result of \citet{Yoon19} adopting a linear galaxy bias model, which may not be a sufficient assumption to interpret the data points to scales as small as $\ell \sim 2000$ (see Fig.~7 of \citealt{Krause17} for the determination of scale cuts in DES Y1 galaxy lensing and galaxy clustering observables to avoid the impact of non-linear galaxy bias).

Although baryonic effects are the dominant systematic uncertainty on small-scale cosmic shear measurements, there are other systematics that will likely become important in future, more constraining analyses. 
Contributions from third order corrections of the shear two-point correlations, such as reduced shear \citep{Shapiro09} and magnification bias \citep{Schmidt09} effects are estimated to produce a $\sim 2\%$ fractional difference in the observables of $\xi_+$ and $\sim 5\%$ in $\xi_-$ at 2.5$\arcmin$. If not accounted for, they would lead to a $\sim 1\sigma$-level bias in the constraint of the \texttt{HMcode} baryon parameter \citep{Mead15} under a DES Y5-like data quality, according to \citet{MacCrann17} (see their Fig.~5 and Fig.~7). 
The choice of IA models can also affect baryon constraints in future more constraining data sets, given that IA and cosmological parameters are degenerate (see, e.g., Fig.~4 of \citealt{MacCrann17}). For Y1, switching from the simple NLA model to the full tidal alignment and tidal torquing (TATT) model \citep{Blazek19} leads to a slight shift of $\sim 0.5\sigma$ in $S_8$, but overall the resulting likelihoods are still in agreement within Y1 errors \citep{Troxel18, Samuroff18}. 
The improved data quality in forthcoming datasets will likely mean that discrepancies induced from these small-scale systematics will become non-negligible, and will require extra efforts to extract precise joint constrains on both cosmology and baryonic physics.

The ongoing KiDS and HSC analyses provide an excellent dataset to get additional insights into discriminating between different baryonic physics scenarios. Moreover, future datasets from DES Year 3 and Year 6 will provide improved joint constraints on baryonic physics and cosmological parameters.

 In the regime where effects of baryonic physics are causing the suppression of clustering power ($k \lesssim$ few $10$ Mpc$^{-1}h$), the properties of halo gas contain a wealth of information on baryon feedback mechanisms. Observational probes such as X-ray, thermal and kinetic Sunyaev-Zel'dovich measurements are directly sensitive to the distribution and the characteristics of gas content (e.g. \citealt{Battaglia17}). Ultimately, utilizing information from both gas-sensitive observables and weak lensing provides the most promising avenue to constrain baryon feedback \citep{ Hojjati17, Pandey20, Osato20, Debackere20, Arico20, Mead20, Schneider20b}. 

As we prepare for future analyses of the Rubin Observatory LSST, Euclid, SPHEREx, and Roman Space Telescope, the information regarding which baryonic scenarios are already excluded by Stage III data is invaluable for the design of cosmology analysis pipelines and simulation efforts in order to focus the computational power where it is needed most and in order to optimally analyze these future data sets.

\section*{Acknowledgements}
We thank the internal reviewers from the DES collaboration for providing insightful comments and giving valuable feedback. HH and TE are supported by NASA ROSES ADAP, grant 16-ADAP16-0116 and Department of Energy Cosmic Frontier program, grant DE-SC0020215. RM is supported by the Department of Energy Cosmic Frontier program, grant DE-SC0010118.

\section*{Data Availability Statement}
The observational data underlying this article are publicly available in the official DES Year 1 key project website at \href{https://des.ncsa.illinois.edu/releases/y1a1/key-products}{https://des.ncsa.illinois.edu/releases/y1a1/key-products}.



\appendix
\section{The interplay between baryons, cosmology, and other systematic parameters}
\label{sec:parameter_degeneracy}

In this appendix, we investigate the degeneracies of the baryonic physics parameter $Q_1$ with cosmological and other nuisance parameters (see Table~\ref{tb:params}).

\begin{figure*}
  \begin{center}
  \includegraphics[width=0.96\textwidth]{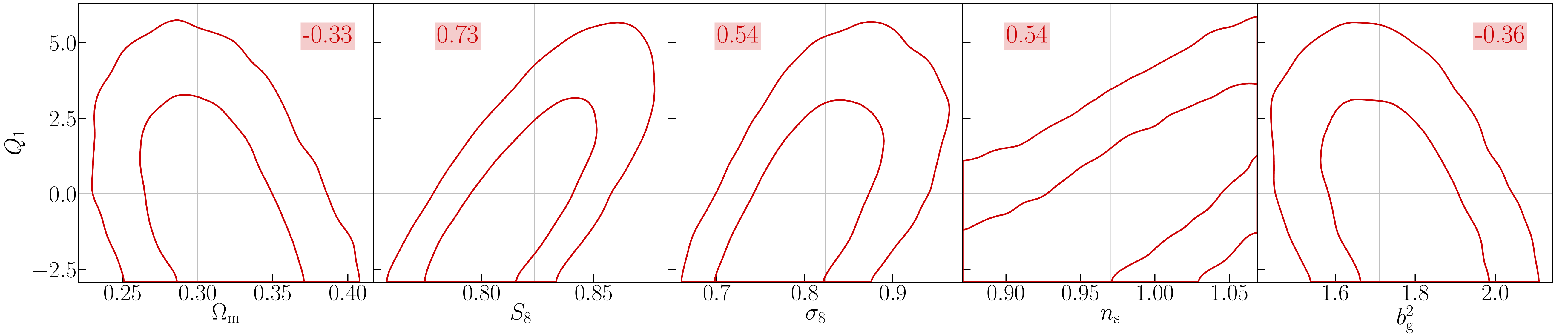}
  \caption{Marginalized 2D posterior distributions between $Q_1$ with parameters that show apparent parameter degeneracies, from the likelihood simulated analysis with the DMO mock data vector as input. The fiducial parameter values are indicated by the cross gray lines. The correlation coefficients are provided in the legend.} 
  \label{fig:parameter_degeneracy}
  \end{center}
\end{figure*}

To quantify the level of parameter degeneracies, we compute correlation coefficients of the marginalized 2D posterior distributions between $Q_1$ with all the other parameters. 
The parameter correlation ${\bf Corr}_{\rm par}^{ij}$ is computed via:
\begin{equation}
  \label{eq:Corr_par}
  {\bf Corr}_{\rm par}^{\rm ij} = \C^{ij}_{\rm par} / \sqrt{\C^{ii}_{\rm par} \C^{jj}_{\rm par}} \ , 
\end{equation} with the parameter covariance matrix computed as
\begin{equation}
\label{eq:Cov_par}
\C^{ij}_{\rm par} = \frac{1}{N-1} \sum^{N}_{k=1} (\bm \theta^{ik}-\langle \bm \theta^i\rangle)( \bm \theta^{jk}-\langle \bm \theta^j\rangle)\ .
\end{equation} 
The $\langle \bm \theta^i\rangle$ indicates the mean of the $i$-th parameter, and $k \in [1,N]$ is the index running over the first 90\% higher likelihood steps in the MCMC chain. We discard the 10\% of the MCMC samples with the lowest likelihood values when deriving the parameter covariance, in order to decrease the effects from samples distributed far away from the high likelihood region.

Using the likelihood simulation chain with the DMO scenario as mock data, in Fig.~\ref{fig:parameter_degeneracy} we display the posterior distributions between $Q_1$ and parameters that are significantly correlated with it.

\begin{figure*}
  \begin{center}
  \includegraphics[width=0.99\textwidth]{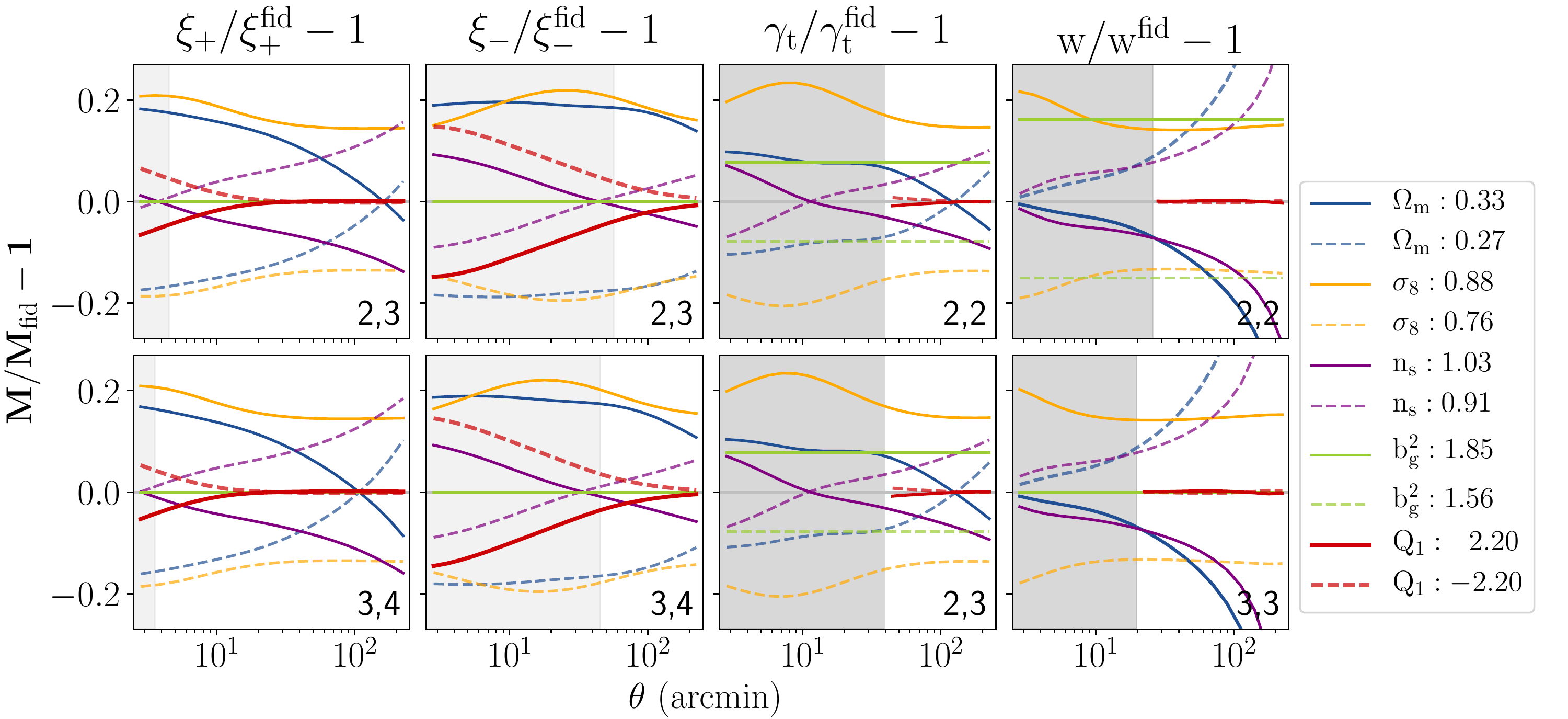}
  \caption{The fractional changes in model vectors when varying each of the individual cosmological or baryonic parameters to $1\sigma$ above (solid lines) or below (dash lines) their fiducial values listed in Table~\ref{tb:params}. We only select two tomographic bins for each of the observables as demonstration, with the bin information indicated on the bottom right corner of each panel. For galaxy-galaxy lensing, the first number is for lens tomographic bin; the second number for source. 
  The darker gray bands in galaxy-galaxy lensing and clustering panels mark data points that are excluded throughout this work. The lighter gray bands in the cosmic shear panels highlight data points that are excluded in the original DES Y1 analysis, but are now included in this work. When varying cosmological parameters, we carefully adjust the $A_s$ parameter in \textsc{CosmoLike} in order to keep $\sigma_8$ fixed.} 
  \label{fig:Mpar_1sigma_sym}
  \end{center}
\end{figure*}

To understand the trends of parameter degeneracies from the MCMC, in Fig.~\ref{fig:Mpar_1sigma_sym}, we plot the fractional changes in model data vectors ($\frac{\M-\M_{\rm fid}}{\M_{\rm fid}}$) when varying individual parameters to $1\sigma$ above (solid lines) or below (dash lines) from their fiducial values listed in Table~\ref{tb:params}. 
The positive correlation trend between $Q_1$ and $n_{\rm s}$ is clearly due to their opposing effects on the model vector, especially in $\xi_{-}$, as shown in the second column of Fig.~\ref{fig:Mpar_1sigma_sym} (red vs.\ purple curves).

The significant positive correlation between $Q_1$ and $S_8$ explains the tendency towards parameter projection effects discussed in \S\ref{subsec:NPCmodes_determination} and \S\ref{subsec:baryon_constraint_simulated}. 
This correlation is straightforward to understand. An increase in $S_8$ boosts the overall amplitude of matter clustering, whereas increasing the amount of feedback suppresses the clustering signal on small scales. 
The opposite correlation directions between $Q_1$ with $\Omegam$ (-0.33) and with $\sigma_8$ (0.54) are driven by the significant negative coupling between $\Omegam$ and $\sigma_8$. 

Regarding the negative correlation observed between $Q_1$ and the galaxy bias parameters\footnote{In Fig.~\ref{fig:parameter_degeneracy} we pick the galaxy bias parameter of the second tomographic bin $b^{2}_{\rm g}$ as a demonstration; however, the parameter correlations between $Q_1$ and other tomographic bins $b^{i}_{\rm g}$ show similar results.}, this correlation is actually driven by the common degeneracies of $Q_1$ and $b^{i}_{\rm g}$ with $S_8$. 
Ideally there should be almost no correlation between $Q_1$ and $b^{i}_{\rm g}$ because the variation of $Q_1$ is mostly affecting the cosmic shear observables, whereas the galaxy bias parameters only affect galaxy-galaxy lensing and galaxy clustering (see Fig.~\ref{fig:Mpar_1sigma_sym}). However, when cosmology is allowed to vary, $Q_1$ and $b^{i}_{\rm g}$ appear correlated because of their correlation with cosmological parameters (mostly in $\Omegam$ and $\sigma_8$).\footnote{The same logic also explains why there are tight correlations between pairs of bias parameters ($b^{i}_{\rm g}$ and $b^{j}_{\rm g}$). In principle, each bias parameter governs different portions of the model vector via the tomography bin division (if there are no photo-$z$ uncertainties when dividing lens galaxies), and thus should have no correlation. The observed high correlations among the bias parameters are driven by the fact that they have a similar impact on $\Omegam$ and $\sigma_8$.}




\bibliographystyle{mnras}
\bibliography{reference} 

\section*{Author Affiliations}
\input{author_affiliation.tex}

\bsp	
\label{lastpage}
\end{document}

%% file: def.tex
\newcommand{\U}{\mathbf U}
\newcommand{\V}{\mathbf V}
\renewcommand{\t}{\rm t}
\newcommand{\D}{\bm D}
\newcommand{\M}{\bm M}
\newcommand{\B}{\bm B}

\newcommand{\Q}{\bm Q}
\renewcommand{\P}{\mathbf P}
\newcommand{\C}{\mathbf C}

\newcommand{\invL}{\mathbf L^{-1}}
\renewcommand{\L}{\mathbf L}
\newcommand{\pco}{\bm p_{\rm co}}
\newcommand{\pnu}{\bm p_{\rm nu}}
\newcommand{\pcosim}{\bm p_{\rm co, sim}}

\newcommand{\Msun}{\mathrm{M}_{\odot}}

\newcommand{\metacal}{{\textsc{metacalibration} }} 

\newcommand\AIA{A_{\rm IA}}
\newcommand\etaIA{\eta_{\rm IA}}
\newcommand{\photoz}{photo-$z$}

\newcommand{\LCDM}{$\Lambda$CDM }
\newcommand{\Omegam}{\Omega_{\mathrm{m}}}
\newcommand{\Omegab}{\Omega_{\mathrm{b}}}

\newcommand{\ns}{n_{\mathrm{s}}}

%% file: author_list.tex
\author[Huang et al.]{
\parbox{\textwidth}{
\Large
Hung-Jin Huang,$^{1, 2}$\thanks{E-mail: hungjinh@email.arizona.edu}
Tim Eifler,$^{1}$
Rachel Mandelbaum,$^{2}$
Gary M. Bernstein,$^{3}$
Anqi Chen,$^{4}$
Ami Choi,$^{5}$
Juan Garc\'ia-Bellido,$^{6}$
Dragan Huterer,$^{4}$
Elisabeth Krause,$^{1}$
Eduardo Rozo,$^{7}$
Sukhdeep Singh,$^{8}$
Sarah Bridle,$^{9}$
Joseph DeRose,$^{8, 10}$
Jack Elvin-Poole,$^{5, 11}$
Xiao Fang,$^{1}$
Oliver Friedrich,$^{12}$
Marco Gatti,$^{13}$
Enrique Gaztanaga,$^{14, 15}$
Daniel Gruen,$^{16, 17, 18}$
Will Hartley,$^{19, 20, 21}$
Ben Hoyle,$^{22, 23}$
Mike Jarvis,$^{3}$
Niall MacCrann,$^{5, 11}$
Vivian Miranda,$^{1}$
Markus Rau,$^{2}$
Judit Prat,$^{24}$
Carles S\'anchez,$^{3}$
Simon Samuroff,$^{2}$
Michael Troxel,$^{25}$
Joe Zuntz,$^{26}$
Tim Abbott,$^{27}$
Michel Aguena,$^{28, 29}$
James Annis,$^{30}$
Santiago Avila,$^{6}$
Matthew Becker,$^{31}$
Emmanuel Bertin,$^{32, 33}$
David Brooks,$^{20}$
David Burke,$^{17, 18}$
Aurelio Carnero Rosell,$^{34, 35}$
Matias Carrasco Kind,$^{36, 37}$
Jorge Carretero,$^{13}$
Francisco Javier Castander,$^{14, 15}$
Luiz da Costa,$^{29, 38}$
Juan De Vicente,$^{39}$
J\"org Dietrich,$^{40}$
Peter Doel,$^{20}$
Spencer Everett,$^{10}$
Brenna Flaugher,$^{30}$
Pablo Fosalba,$^{14, 15}$
Josh Frieman,$^{30, 41}$
Robert Gruendl,$^{36, 37}$
Gaston Gutierrez,$^{30}$
Samuel Hinton,$^{42}$
Klaus Honscheid,$^{5, 11}$
David James,$^{43}$
Kyler Kuehn,$^{44, 45}$
Ofer Lahav,$^{20}$
Marcos Lima,$^{28, 29}$
Marcio Maia,$^{29, 38}$
Jennifer Marshall,$^{46}$
Felipe Menanteau,$^{36, 37}$
Ramon Miquel,$^{13, 47}$
Francisco Paz-Chinch\'{o}n,$^{12, 37}$ 
Andr\'es Plazas Malag\'on,$^{48}$
Kathy Romer,$^{49}$
Aaron Roodman,$^{17, 18}$
Eusebio Sanchez,$^{39}$
Vic Scarpine,$^{30}$
Santiago Serrano,$^{14, 15}$
Ignacio Sevilla,$^{39}$
Mathew Smith,$^{50}$
Marcelle Soares-Santos,$^{51}$
Eric Suchyta,$^{52}$
Molly Swanson,$^{37}$
Gregory Tarle,$^{4}$
Diehl H. Thomas,$^{30}$
Jochen Weller$^{22, 23}$ \\
and (The DES Collaboration)
}
\vspace{0.4cm} 
\\
Author affiliations are listed at the end of the paper.
}

%% file: author_affiliation.tex
$^{1}$Steward Observatory/Department of Astronomy, University of Arizona, 933 North Cherry Avenue, Tucson, AZ 85721, USA \\
$^{2}$McWilliams Center for Cosmology, Department of Physics, Carnegie Mellon University, Pittsburgh, PA 15213, USA \\
$^{3}$Department of Physics and Astronomy, University of Pennsylvania, Philadelphia, PA 19104, USA \\
$^{4}$Department of Physics, University of Michigan, Ann Arbor, MI 48109, USA \\
$^{5}$Center for Cosmology and Astro-Particle Physics, The Ohio State University, Columbus, OH 43210, USA\\
$^{6}$Instituto de Fisica Teorica UAM/CSIC, Universidad Autonoma de Madrid, 28049 Madrid, Spain\\
$^{7}$Department of Physics, University of Arizona, Tucson, AZ 85721, USA\\
$^{8}$Berkeley Center for Cosmological Physics, University of California, Berkeley, CA 94720, USA\\
$^{9}$Jodrell Bank Center for Astrophysics, School of Physics and Astronomy, University of Manchester, Oxford Road, Manchester, M13 9PL, UK \\
$^{10}$ Santa Cruz Institute for Particle Physics, Santa Cruz, CA 95064, USA\\
$^{11}$ Department of Physics, The Ohio State University, Columbus, OH 43210, USA\\
$^{12}$ Kavli Institute for Cosmology, University of Cambridge, Madingley Road, Cambridge CB3 0HA, UK\\
$^{13}$ Institut de F\'{\i}sica d'Altes Energies (IFAE), The Barcelona Institute of Science and Technology, Campus UAB, 08193 Bellaterra (Barcelona) Spain\\
$^{14}$ Institut d'Estudis Espacials de Catalunya (IEEC), 08034 Barcelona, Spain\\
$^{15}$ Institute of Space Sciences (ICE, CSIC),  Campus UAB, Carrer de Can Magrans, s/n,  08193 Barcelona, Spain\\
$^{16}$ Department of Physics, Stanford University, 382 Via Pueblo Mall, Stanford, CA 94305, USA\\
$^{17}$ Kavli Institute for Particle Astrophysics \& Cosmology, P. O. Box 2450, Stanford University, Stanford, CA 94305, USA\\
$^{18}$ SLAC National Accelerator Laboratory, Menlo Park, CA 94025, USA\\
$^{19}$ D\'{e}partement de Physique Th\'{e}orique and Center for Astroparticle Physics, Universit\'{e} de Gen\`{e}ve, 24 quai Ernest Ansermet, CH-1211 Geneva, Switzerland\\
$^{20}$ Department of Physics \& Astronomy, University College London, Gower Street, London, WC1E 6BT, UK\\
$^{21}$ Department of Physics, ETH Zurich, Wolfgang-Pauli-Strasse 16, CH-8093 Zurich, Switzerland\\
$^{22}$ Max Planck Institute for Extraterrestrial Physics, Giessenbachstrasse, 85748 Garching, Germany\\
$^{23}$ Universit\""ats-Sternwarte, Fakult\""at f\""ur Physik, Ludwig-Maximilians Universit\""at M\""unchen, Scheinerstr. 1, 81679 M\""unchen, Germany\\
$^{24}$ Department of Astronomy and Astrophysics, University of Chicago, Chicago, IL 60637, USA\\
$^{25}$ Department of Physics, Duke University Durham, NC 27708, USA\\
$^{26}$ Institute for Astronomy, University of Edinburgh, Edinburgh EH9 3HJ, UK\\
$^{27}$ Cerro Tololo Inter-American Observatory, NSF's National Optical-Infrared Astronomy Research Laboratory, Casilla 603, La Serena, Chile\\
$^{28}$ Departamento de F\'isica Matem\'atica, Instituto de F\'isica, Universidade de S\~ao Paulo, CP 66318, S\~ao Paulo, SP, 05314-970, Brazil\\
$^{29}$ Laborat\'orio Interinstitucional de e-Astronomia - LIneA, Rua Gal. Jos\'e Cristino 77, Rio de Janeiro, RJ - 20921-400, Brazil\\
$^{30}$ Fermi National Accelerator Laboratory, P. O. Box 500, Batavia, IL 60510, USA\\
$^{31}$ Argonne National Laboratory, 9700 South Cass Avenue, Lemont, IL 60439, USA\\
$^{32}$ CNRS, UMR 7095, Institut d'Astrophysique de Paris, F-75014, Paris, France\\
$^{33}$ Sorbonne Universit\'es, UPMC Univ Paris 06, UMR 7095, Institut d'Astrophysique de Paris, F-75014, Paris, France\\
$^{34}$ Instituto de Astrofisica de Canarias, E-38205 La Laguna, Tenerife, Spain\\
$^{35}$ Universidad de La Laguna, Dpto. AstrofÃ­sica, E-38206 La Laguna, Tenerife, Spain\\
$^{36}$ Department of Astronomy, University of Illinois at Urbana-Champaign, 1002 W. Green Street, Urbana, IL 61801, USA\\
$^{37}$ National Center for Supercomputing Applications, 1205 West Clark St., Urbana, IL 61801, USA\\
$^{38}$ Observat\'orio Nacional, Rua Gal. Jos\'e Cristino 77, Rio de Janeiro, RJ - 20921-400, Brazil\\
$^{39}$ Centro de Investigaciones Energ\'eticas, Medioambientales y Tecnol\'ogicas (CIEMAT), Madrid, Spain\\
$^{40}$ Faculty of Physics, Ludwig-Maximilians-Universit\""at, Scheinerstr. 1, 81679 Munich, Germany\\
$^{41}$ Kavli Institute for Cosmological Physics, University of Chicago, Chicago, IL 60637, USA\\
$^{42}$ School of Mathematics and Physics, University of Queensland,  Brisbane, QLD 4072, Australia\\
$^{43}$ Center for Astrophysics $\vert$ Harvard \& Smithsonian, 60 Garden Street, Cambridge, MA 02138, USA\\
$^{44}$ Australian Astronomical Optics, Macquarie University, North Ryde, NSW 2113, Australia\\
$^{45}$ Lowell Observatory, 1400 Mars Hill Rd, Flagstaff, AZ 86001, USA\\
$^{46}$ George P. and Cynthia Woods Mitchell Institute for Fundamental Physics and Astronomy, and Department of Physics and Astronomy, Texas A\&M University, College Station, TX 77843,  USA\\
$^{47}$ Instituci\'o Catalana de Recerca i Estudis Avan\c{c}ats, E-08010 Barcelona, Spain\\
$^{48}$ Department of Astrophysical Sciences, Princeton University, Peyton Hall, Princeton, NJ 08544, USA\\
$^{49}$ Department of Physics and Astronomy, Pevensey Building, University of Sussex, Brighton, BN1 9QH, UK\\
$^{50}$ School of Physics and Astronomy, University of Southampton,  Southampton, SO17 1BJ, UK\\
$^{51}$ Brandeis University, Physics Department, 415 South Street, Waltham MA 02453\\
$^{52}$ Computer Science and Mathematics Division, Oak Ridge National Laboratory, Oak Ridge, TN 37831\\